\documentclass{article}

\usepackage{amssymb,amsfonts,amsmath}
\usepackage{cite,enumerate,float,indentfirst}
\usepackage{color}

\def\be{\begin{eqnarray}}
\def\ee{\end{eqnarray}}
\def\nn{\nonumber}

\def\tr{{\rm tr}\,}
\def\Tr{{\rm Tr}\,}

\def\bp{\underline{{\bf p}}}

\def\Schur{{\rm Schur}}
\def\Mac{{\rm Mac}}
\def\Ker{{\rm Ker}}

\definecolor{red}{rgb}{1,0,0}
\definecolor{orange}{rgb}{1,0.5,0}
\definecolor{violet}{rgb}{0.7,0,1}

\hoffset= - 1.0in         % switch off for draft style
\begin{document}

\title{\vspace{-.5cm}{\Large {\bf  Kerov functions for composite representations and Macdonald ideal}\vspace{-.5cm}}
\author{
{\bf A.Mironov$^{a,b,c}$}\footnote{mironov@lpi.ru; mironov@itep.ru}\ \ and
\ {\bf A.Morozov$^{b,c}$}\thanks{morozov@itep.ru}}
\date{ }
}

\maketitle

\vspace{-5cm}

\begin{center}
\hfill FIAN/TD-01/19\\
\hfill IITP/TH-03/19\\
\hfill ITEP/TH-03/19
\end{center}

\vspace{3cm}

\begin{center}
$^a$ {\small {\it Lebedev Physics Institute, Moscow 119991, Russia}}\\
$^b$ {\small {\it ITEP, Moscow 117218, Russia}}\\
$^c$ {\small {\it Institute for Information Transmission Problems, Moscow 127994, Russia}}

\end{center}

\vspace{.5cm}

\begin{abstract}
Kerov functions provide an infinite-parametric deformation of the set of Schur functions,
which is a far-going generalization of the 2-parametric Macdonald deformation.
In this paper, we concentrate on a particular subject:
on Kerov functions labeled by the Young diagrams associated with the conjugate and, more generally, composite representations.
Our description highlights peculiarities of the Macdonald locus (ideal)
in the space of the Kerov parameters,
where {\it some} formulas and relations get drastically simplified.
However, even in this case, they substantially deviate from the Schur case,
which illustrates
the problems encountered in the theory of link hyperpolynomials.
An important additional feature of the Macdonald case is {\it uniformization},
a possibility of capturing the dependence on $N$ for symmetric polynomials of $N$
variables into a single variable $A=t^N$, while in the generic Kerov case the
$N$-dependence looks considerably more involved.
\end{abstract}

\vspace{.0cm}

\section{Introduction}

In this paper, we continue our study of theory of Kerov functions\footnote{The Kerov {\it function} should not be confused
with much better known  "Kerov character polynomials",
also associated with the name of S.Kerov \cite{Kerpols}:
to avoid the confusion, we use the term {\it functions}
throughout the present text.} \cite{Kerov,FZ,Kerovmore}
and their applications.
We give a brief summary of the issues already reviewed in \cite{MMker}
and proceed to the very important set of questions related to the Kerov functions labeled by the Young diagrams associated with the conjugate and, more generally, composite representations. Though, in the Kerov case, literal representation theory does not work, for the sake of simplicity, we refer to these Kerov functions as associated with composite representation.
They are important in all applications, especially in the theory
of topological vertices.
As we shall see, simplicity of this story in the non-refined case, i.e. for the Schur polynomials, was actually
due to a very simple conjugation rule
\be
{\rm Schur}_{(0,P)}[X] := {\rm Schur}_{\bar P}[X] = \Big(\det X\Big)^{P_1}\cdot \Schur_{P}[X^{-1}]
\label{conjSchur}
\ee
which has a natural lifting to composite representations in the form
of the Koike formula \cite{Koike,Kanno,MMhopf},
\be
{\rm \Schur}_{(R,P)}[X] =  \Big(\det X\Big)^{P_1}\cdot
\sum_\eta (-1)^{|\eta|}\Schur_{R/\eta}[X]\cdot\Schur_{P/\eta^\vee}[X^{-1}]
\label{compSchur}
\ee
Here $X$ is the $N\times N$ matrix defining the $N$-dimensional Miwa locus in the space
of ordinary time variables $p_k = \tr X^k$, $R$ and $P$ are the Young diagrams, $P_1\ge P_2\ge\ldots\ge P_{l_{_P}}\ge 0$ and similarly for $R$, and $\bar P$ is the conjugate of $P$. The Schur polynomials are symmetric functions of the eigenvalues $x_a$ of the matrix $X$ \cite{Mac,Fulton}.

Our goal is to describe a Kerov version of these formulas,
and the result is that the sum at the r.h.s. of (\ref{compSchur})
extends to a double sum over all the subdiagrams,
and generically this is true not only for the Kerov functions, but also for
the Macdonald polynomials \cite{Mac}, though the formulas in the latter case are much simpler.
These extra terms cause problems for the refined version of \cite{L8n8},
i.e. for constructing non-torus hyperpolynomials
for the simplest link $L_{8n8}$ from conventional
refined topological vertex of \cite{refTV}.

More than that, even the Kerov counterpart of (\ref{conjSchur}) acquires a full-fledged sum
at the r.h.s. over all the Young diagrams of the size (number of boxes) $|R|$.
In variance with (\ref{compSchur}), this effect {\it disappears}
on the Macdonald locus, and this is an example of the situation,
where the Macdonald polynomials are simpler than the generic Kerov ones.
However, after a minor change of the question: how to get from (\ref{conjSchur})
to (\ref{compSchur}) the simplicity disappears, at least partly.
Of course, simpler-looking are also particular coefficients in
the Macdonald deformation of (\ref{compSchur}).
However, this may depend on the choice of notation:
it is not very clear what kind of {\it properties} of these coefficients
is truly simplified.

We describe only one of such properties in addition to above-mentioned
nullification of some coefficients,
the {\it uniformization} \cite{univ}, i.e. a possibility of
capturing the $N$-dependence into a single parameter $A$,
on which the coefficients depend rationally like on all other parameters.
As explained in \cite{MMker},
the Kerov functions can be considered as
depending on two sets of variables, which in many respects are similar:
$\Ker_R^{(g)}\{p_k\} = \Ker_R\{p_k|g_k\}$.
The Macdonald locus is given by the relations $g_k=\frac{\{q^k\}}{\{t^k\}}$, where, hereafter, we use the notation $\{x\}:=x-x^{-1}$, and plays for the $g$-variables
the same role as the topological locus $p_k=\frac{\{a^k\}}{\{t^k\}}$
does for the $p$-variables, thus one can expect various factorizations
to occur, and this is what really happens and leads to {\it uniformization}
of the coefficients.
However, an explicit description of $p-g$ duality is still out of reach,
and a systematic conversion of $p$-properties into $g$-properties is
not yet available.

Of course, all this requires  re-thinking and insights,
we just make a first step in this direction and do not pretend  to put
the right accents on various observations we make.
We begin in sec.2 with reminding different issues relevant for
our discussion.
The two auxiliary sections 3 and 4 are devoted to the two important aspects
of the generalizations: to denominator functions and to description
of Macdonald locus as an ideal in the space of $g$-polynomials.
Then sec.5 and 6 discuss Kerov and Macdonald generalizations of the Koike formulas
(\ref{conjSchur}) and (\ref{compSchur}).
All these results, summarized in sec.7 are largely speculative but already challenging.
They call for a deep study and understanding.

\section{Definitions and summaries}

\subsection{Kerov functions. Summary}

\subsubsection{Definitions}

Kerov functions are symmetric functions (polynomials) of variables $x_a$ (eigenvalues of a matrix $X$), or, equivalently, of time variables
$p_k:=\sum_a x_a^k=\tr X^k$, hence they are labeled by Young diagrams, the degree of the polynomial being size of the Young diagram $|R|$. They are also rational functions of the infinite set of parameters $g_k$ and depend on the first $|R|$ of them.

Following \cite{Kerov} and \cite{MMker}, we define a pair of Kerov functions as triangular combinations of Schur functions in two different orderings of Young diagrams,
\be
\left\{
\begin{array}{c}
\Ker^{(g)}_R\{ p\} =  {\rm Schur}_{R}\{p\}
+ \sum_{{R'< R}} {\cal K}_{{R,R'}}^{(g)} \cdot {\rm Schur}_{R'}\{p\}
\\ \\
\widehat{\Ker}_R^{(g )}\{p\}
=\Schur_{R}\{p\}\ + \  \sum_{ R'^\vee> R^\vee}
\widehat{\cal K}^{(g )}_{R'^\vee,R^\vee}\cdot\Schur_{R'}\{p\}
\end{array}
\right.
\label{Kerdef}
\ee
Here we denote $R^\vee$ the transposition of the Young diagram $R$, the sign $<$ refers to the lexicographical ordering,
\be
R>R'  \ \ {\rm if} \ \ r_1>r_1' \ \ {\rm  or\ if} \ \ r_1=r_1', \ {\rm but} \ r_2>r_2',
\ \ {\rm or\  if} \ \ r_1=r_1' \ {\rm  and} \ r_2=r_2', \ {\rm but} \ r_3>r_3',
\ \ {\rm and\ so\ on}
\label{lexico}
\ee
and the two summation rules in (\ref{Kerdef}) begin to deviate from each
other from level $|R|=6$, when there is a pair of Young diagrams,
$[3,1,1,1]>[2,2,2]$, for which $[3,1,1,1]^\vee = [4,1,1]>[2,2,2]^\vee=[3,3]$.

The Macdonald-Kostka coefficients  ${\cal K}_{R,R'}^{(g)}$ in (\ref{Kerdef})
are defined iteratively in $R$ and $R'$
from the orthogonality conditions
\be
\Big<\Ker^{(g)}_R\Big|\Ker^{(g)}_{R'}\Big>
= ||\Ker^{(g)}_R||^2 \cdot \delta_{R,R'}
\label{orthoc}
\ee
w.r.t. the scalar product
\be
\Big< { p}^{\Delta}\Big| { p}^{\Delta'} \Big>^{(g)} =
z_\Delta \cdot \delta_{\Delta,\Delta'}\cdot
\left(\prod_{i=1}^{l_\Delta} g_{\delta_i}\right)
= z_\Delta \cdot g^\Delta\cdot \delta_{\Delta,\Delta'}
\label{scapro}
\ee
i.e. from the Gauss decomposition of the matrix
\be
\mu_{R,R'}=\Big<\Schur_R\Big|\Schur_{R'}\Big>^{(g)}
=\sum_\Delta \ {\psi_{R_1}(\Delta)\psi_{R_2}(\Delta)\over z_\Delta}\cdot g^\Delta
\label{Schurscaprog}
\ee
or any of its powers, positive or negative,
\be
\Big(\mu^{n}\Big)_{R,R'}=\mu^{(g^{n})}_{R,R'}
=\sum_\Delta \ (g^\Delta)^n\cdot {\psi_{R_1}(\Delta)\psi_{R_2}(\Delta)\over z_\Delta}
\ee
Here $\psi_R(\Delta)$ are the characters of symmetric group $S_{|R|}$
and with the Young diagram
$\Delta=\big[\delta_1\geq\delta_2\geq\ldots\geq \delta_{l_\Delta}\big]$,
one associates a monomial $p^\Delta = \prod_{i=1}^{l_\Delta} p_{\delta_i}$. The variables $g_k$ parameterize the measure that defines the Kerov functions.
The combinatorial factor $z_\Delta$ is best defined in the dual parametrization of the
Young diagram, $\Delta = \big[\ldots,2^{m_2},1^{m_1}\big]$, then
$z_\Delta = \prod_k k^{m_k}\cdot m_k!$.
Note that the normalization of $\Ker^{(g)}$ is already fixed by  the choice of
unit diagonal coefficient (the first term) in (\ref{Kerdef}),
${\cal K}^{(g)}_{RR}=1$, so that
the norm $||\Ker^{(g)}||$ is a deducible quantity.

We refer to \cite{MMker} for a comprehensive collection of properties
of Kerov functions (i.e. to the Kerov lifting of {\it all} the relevant
properties of the Schur polynomials).
In the rest of this subsection, we mention some of them in the form
suited for the discussion in the present paper.

\subsubsection{Denominator functions
\label{denof}}

In Kerov calculus, one associate with each Young diagram $R$ the four numbers:
\be
\nu_R = &  \ \hbox{ the sequence number of} \ R \ \text{in the lexicographic ordering}, \nn \\
\nu_R' = &  \ \hbox{ the sequence number of} \ R \ \text{from the end of lexicographic ordering}, \nn \\
\widehat\nu_R = & \ \hbox{ the sequence number of} \ R \ \text{in the transposed lexicographic ordering}, \nn \\
\widehat\nu_R' = &  \ \hbox{ the sequence number of} \ R \
\text{from the end of the transposed lexicographic ordering}
\ee
i.e. $\nu_R'=p(|R|)+1-\nu_R$, $\widehat\nu_R'=p(|R|)+1-\widehat\nu_R$, where $p(n)$ is the number of partitions at level $n$.

Since the Kerov functions are rational functions of $g_k$, the first special functions in Kerov calculus are their
{\it denominators}: $\Delta_R=\Delta_{|R|}^{(\nu_R)}$
and $\hat\Delta_R=\hat\Delta_{|R|}^{(\hat\nu_R)}$.
The shape of these functions is actually controlled by the numbers $\nu$
(see sec.\ref{exaKerthroughDelta} for an explicit example):
\be
\Ker^{(g)}_R\{p\} = \frac{{\rm pol}_R\{p_k,g_k\}}{\Delta_{|R|}^{(\nu_R)}\{g_k\}}
= \frac{{\rm pol}'_R\Big\{\frac{p_k}{g_k},\frac{1}{g_k}\Big\}}
{\hat\Delta_{|R|}^{(\nu_R'+1)}\Big\{\frac{1}{g_k}\Big\}}
\nn \\ \nn \\
\widehat\Ker^{(g)}_R\{p\} = \frac{\widehat{\rm pol}_R\{p_k,g_k\}}
{\hat\Delta_{|R|}^{(\hat\nu_R)}\{g_k\}}
= \frac{\widehat{\rm pol}'_R\Big\{\frac{p_k}{g_k},\frac{1}{g_k}\Big\}}
{\Delta_{|R|}^{(\hat\nu'_R+1)}\Big\{\frac{1}{g_k}\Big\}}
\label{KerthroughDelta}
\ee
Of course, $\hat\Delta_R$ differs from $\Delta_R$ only when $\nu_R\neq \hat\nu_R$.
Denominator functions are positive integer polynomials of $g_k$
(modulo simple factorial multipliers),
the first of them are:
\be
\Delta^{(1)}_r=1 \nn \\ \nn \\
\Delta^{(2)}_r = \Schur_r\{g\} \nn \\ \nn \\
\Delta^{(3)}_{r}
=  \det\left(\begin{array}{ccc}
\Schur_{[r]}\{g\} & 0 & \Schur_{[1]}\{g\} \\
\Schur_{[r]}\{g\} & \Schur_{[r-1]}\{g\} & 0 \\
\Schur_{[r-1]}\{g\}  & \Schur_{[r-2]}\{g\} & 1
\end{array} \right)
\nn\\
\ldots
\label{Delta1to3}
\ee
so that
\be
\nu_{[1^r]}=1 & \Longrightarrow &
\Ker^{(g)}_{[1^r]} = \Schur_{[1^r]}\{p\} \nn \\
\nu_{[2,1^{r-2}]}=2 & \Longrightarrow &
\Ker^{(g)}_{[2,1^{r-2}]} = \frac{\Schur_{[2,1^{r-2}]}\{p\}\cdot\Schur_{[r]}\{g\}
- \Schur_{[1^r]}\{p\}\cdot \Schur_{[r-1,1]}\{g\}}{\Schur_{[r]}\{g\}}
\nn \\
\nu_{[2,2,1^{r-4}]}=3 & \Longrightarrow &
\Ker^{(g)}_{[2,2,1^{r-4}]} = \frac{{\rm pol}_{[2,2,1^{r-4}]}\{p,g\}}{\Delta_r^{(3)}\{g\}}
\nn \\
\ldots
\nn \\
\nu'_{[r-1,1]} = 2 & \Longrightarrow &
\Ker^{(g)}_{[r-1,1]} =
\frac{
\Schur_{[r]}\left\{\frac{1}{g_k}\right\}\cdot\Schur_{[r-1,1]}\left\{\frac{p_k}{g_k}\right\} -
{\Schur_{[r-1,1]}\left\{\frac{1}{g_k}\right\}}\cdot
\Schur_{[r]}\left\{\frac{p_k}{g_k}\right\}}
{\Schur_{[1]}\left\{\frac{1}{g_k}\right\}\cdot\Delta^{(3)}_r\left\{\frac{1}{g_k}\right\}}
\nn \\
\nu'_{[r]} = 1 & \Longrightarrow &
\Ker^{(g)}_{[r]} = \frac{\Schur_{[r]}\Big\{\frac{p_k}{g_k}\Big\}}{\Schur_r\Big\{\frac{1}{g_k}\Big\}}
\ee
As is clear from these examples, the structure of numerators is similar
to the corresponding $\Delta$, but they depend also on $p$-variables and
thus on the shape of $R$ in order to reproduce the Schur functions $\Schur_R\{p\}$
when all $g_k=1$.

The table of ${\rm deg}\Big(\Delta_r^{(\nu)}\Big)$ and ${\rm deg}\Big(\hat\Delta_r^{(\nu)}\Big)$ looks as follows
(when the two numbers do not coincide we indicate ${\rm deg}\Big(\Delta_r^{(\nu)}\Big)/{\rm deg}\Big(\hat\Delta_r^{(\nu)}\Big)$):

{\footnotesize
$$
\begin{array}{c|ccccccccccccccc|}
r\backslash \nu & 1&2 &3 &4&5&6&7 &8
& 9 &10 & 11 &12 &13 & 14 & 15
\\
\hline
1 & 0&&&&&&&&&&&&&& \\
2& 0&2&&&&&&&&&&&&& \\
3 & 0&3&5&&&&&&&&&&&&\\
4 & 0&4&7&11&11&&&&&&&&&&\\
5 & 0&5&9&14&15&19&16&&&&&&&&\\
6 & 0&6&11&17/19&23/17&25&30/28&36/30&34&36&27&&&&\\
7& 0&7&13&20&27/23&30&36&42/35&49/41&48&51&57/43&49&48&34\\
8 & 0&8&15&23&31&39&43&50&57&&&&&&\\
\hline
\end{array}
$$
}

At level 8 the degree is given only for the first set of Kerov functions.

Note that these denominator polynomials have these degrees only upon most of $g_k$ independent: choosing just a few $g_k=\xi^k$ (for an arbitrary $xi$) makes degree of the denominator polynomial lower. E.g. the denominator of $\Ker^{(g)}_{[1,1,2,3]}$ reaches the degree 30 only at most $g_2=g_1^2$. Adding, say, $g_3=g_1^3$ decrease the degree of the denominator. This illustrates a strong correlation between numerators and denominators of the Kerov functions.

\subsubsection{Transposition rule}

It is a straightforward generalization of the transposition rule for the Schur functions:
$\Schur_{R^\vee}\{p\} = (-)^{|R|}\cdot\Schur_R\{-p\}$,
but involves inversion of the parameters $g_k$ and
a switch between the two functions in (\ref{Kerdef}):
\be
\underline{\Ker}_{R }^{(g)}\{p_k\} :=
\frac{\Ker_{R }^{(g)}\{p_k\}}{ ||\Ker_{R }^{(g)}||^2 } =
(-)^{|R|}\cdot
\widehat{\Ker}^{(g^{-1})}_{R^\vee}\Big\{ -\frac{p_k}{g_k} \Big\}
\label{Kertransp}
\ee
In other words, it relates Kerov functions with
$\nu_R = \widehat{\nu}_R'$. Remarkably,
formulas (\ref{KerthroughDelta}) in s.\ref{denof} imply the existence
of a far more interesting relation between the functions
with $\boxed{\widehat{\nu}_R' = \nu_R+1}$, which still remains to
be brought to a simple form like (\ref{Kertransp}).

\subsubsection{Skew Schur functions
and multiplication rule
}

Like the skew Schur polynomials, the skew Kerov functions $\Ker^{(g)}_{R/\eta}\{{  p}\}$ for ordinary representations
$R=(R,\emptyset)$ are defined as functions of arbitrary time-variables
and can be defined
from the decomposition of the functions depending on the sum of time variables:
\be
\Ker_R^{(g)}\{{  p}+{  p'}\} = \sum_{\eta\in R}
\Ker^{(g)}_{R/\eta}\{{  p}\}\cdot \Ker^{(g)}_\eta\{{  p'}\}
\ \ \ \ \ \Longrightarrow \ \ \
\Ker^{(g)}_{R/\zeta}\{{ p}+{\ p'}\} = \sum_{\zeta\in\eta\in R}
\Ker^{(g)}_{R/\eta}\{{  p}\}\cdot \Ker^{(g)}_{\eta/\zeta}\{{  p'}\}
\label{skewviasum}
\ee
They can be obtained by linear combination from the ordinary Kerov functions:
\be
\Ker_{R/\eta}^{(g)}\{{ p}\} = \sum_\zeta \widehat{\bf N}^{R^\vee}_{\eta^\vee\zeta^\vee}(g^{-1})\cdot \Ker_\zeta^{(g)}\{{ p\}}
\label{skewviaHL}
\ee
where  $\widehat{\bf N}$  are the $g$-dependent  Kerov Littlewood-Richardson coefficients, i.e. the structure constants
in the multiplication rule of the {\bf second set} of Kerov functions
\be
\widehat\Ker_{R}^{(g)}\{p\}\cdot \widehat\Ker_{R'}^{(g)}\{p\} = \sum_{\sum_{R^\vee\cup R'^\vee\leq R''^\vee\leq R^\vee + R'^\vee }}
\widehat{\bf N}^{R''}_{RR'}\{g\}\cdot \widehat\Ker^{(g)}_{R''}\{p\}
\label{Kermult}
\ee
and the sums are rather restricted, since the Kerov Littlewood-Richardson coefficients $\widehat{\bf N}^{R''}_{R,R'}(g)$ are non-zero only
in between the partitions $R^\vee\cup R'^\vee = [R^\vee_1+R'^\vee_1,R^\vee_2+R'^\vee_2,\ldots]$ and $R^\vee+R'^\vee = [{\rm ordered\ collection\ of\ all}\
R^\vee_i \ {\rm and} \ R'^\vee_j]$.

Similarly, one can define the second set of skew Kerov functions  $\widehat\Ker^{(g)}_{R/\eta}\{{  p}\}$ with a decomposition formula
like (\ref{skewviasum}). Then, (\ref{skewviaHL}) is replaced with the formula for the second Kerov set, where the structure constants are taken for the first Kerov set, ${\bf N}^{R''}_{RR'}\{g\}$.

Note that the  dual skew Kerov functions,
\be
\underline{\Ker}_R^{(g)}\{p+p'\} =
\sum_{R'\subset R} \underline{\Ker}_{R'}^{(g)}\{p\}
\cdot \underline{\Ker}_{R/R'}^{(g)}\{p'\}
\label{Kersum}
\ee
are decomposed with  ${\bf N}$ themselves:
\be
\underline{\Ker}_{R/R'}^{(g)}\{p\} =
\sum_{R''} {\bf N}^{R}_{{R'} {R''}}(g) \cdot
\underline{\Ker}_{R''}^{(g)}\{p\}
\ee

\subsubsection{Cauchy summation formula}

The Cauchy summation formula remains true for arbitrary Kerov functions
\be
\sum_{R}(-)^{|R|}\cdot\Ker_R^{(g)}\{p\}\cdot\widehat{\Ker}^{(g^{-1})}_{R^\vee}\{-{p'}\}
= \exp\left( \sum_k \frac{p_kp_{k}'}{k}\right)
\label{Cauchy}
\ee
and, more generally,
\be
\sum_{R}(-)^{|R|}\Ker_{R/\eta}^{(g)}\{p\}\cdot
\widehat{\Ker}^{(g^{-1})}_{R^\vee/\zeta}\{-{p'}\}
= \exp\left( \sum_k \frac{p_kp_{k}'}{k}\right)
\cdot
\sum_\sigma (-)^{|\zeta|+|\eta|+|\sigma|}\Ker_{\zeta^\vee/\sigma}^{(g)}\{p\}\cdot
\widehat{\Ker}^{(g^{-1})}_{\eta^\vee/\sigma^\vee}\{-{p'}\}
\ee
where the sum over $\sigma$ at the r.h.s. contains only finitely many terms.

\subsection{Topological locus and its deformations}

\subsubsection{Peculiarities of the Macdonald case}

If one chooses $g_k$ restricted to be $g_k=\frac{\{q^k\}}{\{t^k\}}$, the Kerov functions reduces to the Macdonald polynomials,
which enjoy a series of peculiar properties. The basic ones are:
\begin{itemize}
\item[{\bf (i)}] the Macdonald-Kostka coefficients between the Young diagrams with different sequence numbers in two orderings in (\ref{Kerdef}) is equal to zero;
\item[{\bf (ii)}] some generically non-zero structure constants vanish in the Macdonald case;
\item[{\bf (iii)}] uniformization.
\end{itemize}

Property (i) means that the Macdonald polynomials do not depend on choosing the ordering in (\ref{Kerdef}), i.e. the two sets of Kerov functions coincide on the Macdonald locus $g_k=\frac{\{q^k\}}{\{t^k\}}$. In fact, one can prove that they coincide if and only if $g_k$'s are restricted to the Macdonald locus.

Property (ii) can be important for precise correspondence
with representation theory of finite-dimensional Lie algebras:
for example,  the summation domain in multiplication rule (\ref{Kermult})
is additionally limited by the decomposition rule of the tensor product of representations of group $SL(N)$: $R\otimes R'=\oplus R''$ only on the Macdonald locus,
beyond it all the Kerov Littlewood-Richardson coefficients ${\bf N}$
in (\ref{Kermult}) are non-vanishing. One can again prove that this restriction to the decomposition rule for representations emerges if and only if $g_k$ are put on the Macdonald locus.

The third property (iii), uniformization is a possibility of capturing the dependence of polynomials in conjugate, and, more generally, in composite representations of $SL(N)$ into a single parameter $A$,
on which the coefficients depend rationally like on all other parameters. We will discuss property (iii) in detail in sec.6.

\bigskip

In Kerov theory, a useful look at the Macdonald choice of parameters $g_k$
is to notice its
close relation to a very different subject, the topological locus
for $p$-variables.
Among other things, and together with them
this sheds some light on the factorizations and other apparent simplifications of
many formulas in the Macdonald case.
This also shows the way to understand what happens away of it.

\subsubsection{Topological locus and Macdonald locus}

Since these two loci are going to play a special in this paper,
we remind them once again.

Topological locus (TL) is a specialization of the ordinary time variables
\be
p_k = p_k^*:=\frac{\{A^k\}}{\{t^k\}}
\label{TL}
\ee
and it is the two-dimensional variety in the entire infinite dimensional
space ${\cal P}=\{p_k,k=1,\ldots,\infty\}$ of time variables
where the Schur functions factorize:
\be
\Schur_R^*(A,t) :=\Schur_R\{p_k^*\} = D_R(A,t) = \prod_{i,j\in R}{\{At^{j-i}\}\over \{t^{h_{i,j}}\}}
\ee
where $h_{i,j}$ is length of the hook $(i,j)$.
For $\boxed{A=t^N}$,
this is a well-known hook formula for the quantum dimensions $D_R$
of representation $R$ of $U_t(SL(N))$.
In knot theory, the quantum dimensions are interpreted as values
of the unreduced colored Hopf polynomial for the unknot.

Macdonald locus (ML) is actually the same specialization, but in the
space ${\cal G}$ of Kerov variables $g_k$:
\be
g_k = g_K^{\Mac}:=\frac{\{q^k\}}{\{t^k\}}
\label{ML}
\ee
with  $q$ playing the role of $A$.
Accordingly, the $g$-dependent Schur functions, from which the
Kostka coefficients and other ingredients of Kerov functions are made,
factorize on this locus:
\be
\Schur_Q^{\Mac}(q,t) := \Schur_Q\{g_k^{\Mac}\} = D_Q(q,t) = \prod_{i,j\in R}{\{qt^{j-i}\}\over \{t^{h_{i,j}}\}}
\ee

The Kerov functions depend on two sets of time variables,
$\Ker_R^{(g)}\{p_k\}=\Ker_R\{p_k,g_k\}$,
and when $g_k$ are restricted to the Macdonald locus, they turn into
the Macdonald functions of $p_k$ only, which explicitly depend on $q$ and $t$:
\be
\Mac_R\{p_k,q,t\} := \Ker_R\{p_k,g_k^{\Mac}\}
\ee

An additional non-trivial fact is that, like the Schur polynomials, these are also factorized
on the topological locus, {\bf provided $t$ is the same in (\ref{TL}) and (\ref{ML})},
i.e. the Kerov functions factorize also on the {\it intersection} of these two loci:
a 3-dimensional variety in ${\cal P}\otimes{\cal G}$
which we call Macdonald topological locus (MTL):
\be
\Ker^{**}_R(A,q,t):=\Ker_R\{p_k^*,g_k^{\Mac}\} = \Mac_R\{p_k^*,q,t\} = M^*_R(A,q,t)
= \prod_{i,j\in R}{\{Aq^{j-1}t^{1-i}\}\over \{q^{R_i-j-1}t^{R^\vee_j-i}\}}
\ee
These quantities explicitly described by the above hook formula are often called
Macdonald dimensions, and they provide expressions for
the unreduced colored hyper-polynomials for the unknot.
After a peculiar change of variables $A={\bf a}\sqrt{-{\bf t}}$, $q=-{\bf qt}$, $t={\bf q}$,
which changes sign from minus to plus in the differences (``differentials") $\{Aq^mt^n\}$,
they are interpreted as (unreduced) super-polynomials for the unknot.

\subsubsection{Kerov and Macdonald functions on the Miwa locus $p_k=\Tr X^k$
\label{Miwa}}

Topological locus (\ref{TL}) is a 2-dimensional sub-variety of an $N$-dimensional
Miwa locus $p_k^X=\tr X^k =\sum_{a}^N x_a^k$,
on which the Schur, Macdonald  and Kerov functions
are usually studied in the theory of symmetric functions,
their restrictions to Miwa locus are then denoted through
\be
\boxed{
\Schur_R[X]:=\Schur_R\{p_k^X\},\ \ \ \ \
\Mac_R^{q,t}[X]:=\Mac_R\{p_k^X\}
\ \ \ \ \ {\rm and} \ \ \ \ \
\Ker^{(g)}_R[X]:=\Ker_R\{p_k^X,g_k\}
}
\ee
It plays a very important role in the present paper,
because it is the place where these functions are naturally
defined in $N$-dependent conjugate and composite representations.

Surviving on Miwa locus are only the Schur functions $\Schur_R[X]$ with $l_R\leq N$.
The same remains true for the Macdonald polynomials.
However, this is {\bf not always true} for the first set of Kerov functions:
for example, already at $N=2$  $\Ker_{[r,1,1]}[X]\neq 0$ for $r\geq 4$, at $N=3$ $\Ker_{[r,1,1,1]}[X]\neq 0$ for $r\geq 3$, at $N=4$ $\Ker_{[r,1,1,1,1]}[X]\neq 0$ for $r\geq 3$ etc. Similarly, at $N=2$  $\Ker_{[r,2,1]}[X]\neq 0$ for $r\geq 5$, at $N=3$  $\Ker_{[r,2,1,1]}[X]\neq 0$ for $r\geq 4$, at $N=4$  $\Ker_{[r,2,1,1,1]}[X]\neq 0$ for $r\geq 3$ etc.

The reason for this is that the lexicographic ordering does not imply that
$l_{R'}\geq l_R$ for $R'<R$,
and, therefore, (\ref{Kerdef}) for $\Ker_R$ with $l_R=3$
can include contributions from $\Schur_{R'}$ with $l_{R'}=2$ and, in result,
does not vanish on the Miwa locus for $N=2$.

At the same time, the second ordering implies that $l_{R'}\geq l_R$ for $R'<R$, and
\framebox{$\widehat{\Ker}_R= 0$ whenever $N<l_{_R}$} for the second set of Kerov functions because of the corresponding property of the Schur functions.

\subsubsection{Diagram-dependent deformation of topological locus}

It is sometimes convenient to use a different view on the Miwa locus and a different way to way to introduce it: the discussion of the previous subsection implies a natural association of the Miwa variables with lines of a Young diagram \cite{MMhopf}.
More concretely, we define the deformation of topological locus by a Young diagram $S=[s_1\geq s_2\geq \ldots \geq s_{l_S}>0]$:
\be
{\bf p}^{*S}_k = \left(\prod_i q^{s_i}\right)^{\frac{2k}{N}}\left(\frac{A^k-A^{-k}}{t^k-t^{-k}}
+ A^{-k} \sum_i t^{(2i-1)k}(q^{-2ks_i}-1)\right)
\ee
and
\be
\bp^{*S}_k(A,q,t) = {\bf p}^{*S}_k(A^{-1},q^{-1},t^{-1}) =
\left(\prod_i q^{s_i}\right)^{-\frac{2k}{N}}\left(\frac{A^k-A^{-k}}{t^k-t^{-k}}
+ A^{k} \sum_i t^{-(2i-1)k}(q^{2ks_i}-1)\right)
\label{dialocus}
\ee
In fact, it can be further promoted to
\be
{\bf p}^{*V}_k = \left(\prod_i v_i\right)^{\frac{2k}{N}}\left(\frac{A^k-A^{-k}}{t^k-t^{-k}}
+ A^{-k} \sum_i t^{(2i-1)k}(v_i^{-2k}-1)\right)
\ee
and
\be
\bp^{*V}_k(A,q,t) = {\bf p}^{*V}_k(A^{-1},q^{-1},t^{-1}) =
\left(\prod_i v_i\right)^{-\frac{2k}{N}}\left(\frac{A^k-A^{-k}}{t^k-t^{-k}}
+ A^{k} \sum_i t^{-(2i-1)k}(v_i^{2k}-1)\right)
\ee
since $v_i$ do not need to be made from exponentials of the ordered integers $q^{s_i}$.
Lifting (\ref{dialocus}) to composite representations $S\longrightarrow (R,P)$
is also straightforward:
\be
{\bf p}^{*(R,P)}_k=
 \left(\prod_{i,j} q^{R_i-P_j}\right)^{\frac{2k}{N}}\cdot\left(
{A^k-A^{-k}\over t^k-t^{-k}}+A^{-k}\sum_i t^{(2i-1)k}\Big(q^{-2kR_i}-1\Big)
+A^{k}\sum_i t^{(1-2i)k}\Big(q^{2kP_i}-1\Big)\right)
\label{p*comp}
\ee
The factors in all these definitions correspond to using the Miwa variable formalism with
\be
{\bf p}^{*x}_k  =
\left(\prod_{a=1}^N x_a\right)^{-\frac{k}{N}}\cdot \sum_{a=1}^N x_a^k, \ \ \ \ \ \
\bp^{*x}_k  = \left(\prod_{a=1}^N x_a\right)^{\frac{k}{N}}
\cdot\sum_{a=1}^N x_a^{-k}
\ee

The meaning of this deformed topological locus is not that immediate,
however, it is the central ingredient of the character realization
for the Hopf HOMFLY-PT polynomial \cite{knotpol}:
the topological peculiarity of the Hopf link \cite{Hopf,MMhopf} implies that
\be
H^{\rm Hopf}_{(R_1\otimes R_2)\times S} = H^{\rm Hopf}_{R_1\times S} \cdot
H^{\rm Hopf}_{R_2\times S} \ \ \Longrightarrow \ \
H^{\rm Hopf}_{R\times S} \sim {\rm Schur}_R
\ee
and ${\bf p}^{*S}$ at $t=q$ is the argument of this Schur polynomial,
\be
H^{\rm Hopf}_{R\times S} = D_Q\cdot  {\rm Schur}_R \{{\bf p}^{*S}_{t=q}\}
\ee
Then (\ref{p*comp}) appears in the description of composite representations $(R,P)$.
See \cite{MMhopf} for detailed discussion and references.

Surprisingly or not, the $(q,t)$-deformation of deformed locus is straightforward.

\subsection{Composite representations\label{Comp}}

The composite representation is the most general finite-dimensional irreducible highest weight representations of $SL(N)$  \cite{Koike,GW,MarK}, which are associated with the $N$-dependent
Young diagram
$$(R,S)= \Big[r_1+s_1,\ldots,r_{l_R}+s_1,\underbrace{s_1,\ldots,s_1}_{N-l_{\!_R}-l_{\!_S}},
s_1-s_{_{l_{\!_S}}},p_1-p_{{l_{\!_S}-1}},\ldots,s_1-s_2\Big]
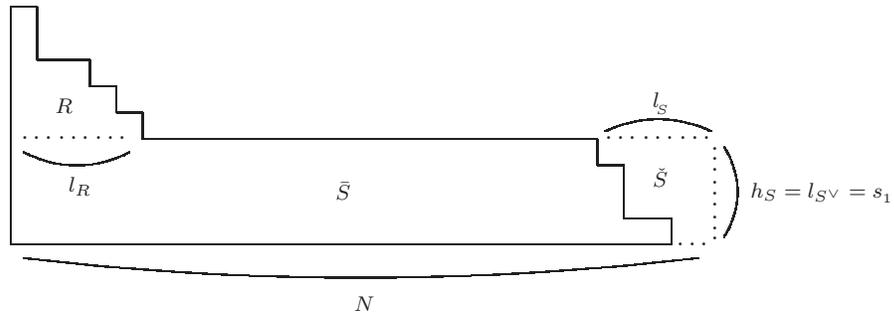
$$

\begin{picture}(300,125)(-90,-30)

\put(0,0){\line(0,1){90}}
\put(0,0){\line(1,0){250}}
\put(50,40){\line(1,0){172}}
%\put(250,0){\line(0,1){40}}

\put(0,90){\line(1,0){10}}
\put(10,90){\line(0,-1){20}}
\put(10,70){\line(1,0){20}}
\put(30,70){\line(0,-1){10}}
\put(30,60){\line(1,0){10}}
\put(40,60){\line(0,-1){10}}
\put(40,50){\line(1,0){10}}
\put(50,50){\line(0,-1){10}}

\put(265,2){\mbox{$\vdots$}}
\put(265,15){\mbox{$\vdots$}}
\put(265,28){\mbox{$\vdots$}}

\put(252,0){\mbox{$\ldots$}}
\put(253,40){\mbox{$\ldots$}}
\put(239,40){\mbox{$\ldots$}}
\put(225,40){\mbox{$\ldots$}}

\put(222,40){\line(0,-1){10}}
\put(222,30){\line(1,0){10}}
\put(232,30){\line(0,-1){20}}
\put(232,10){\line(1,0){18}}
\put(250,0){\line(0,1){10}}

\put(0,90){\line(1,0){10}}
\put(10,90){\line(0,-1){20}}
\put(10,70){\line(1,0){20}}
\put(30,70){\line(0,-1){10}}
\put(30,60){\line(1,0){10}}
\put(40,60){\line(0,-1){10}}
\put(40,50){\line(1,0){10}}
\put(50,50){\line(0,-1){10}}

%\put(-60,40){\mbox{$(R,P) \ \ =$}}
%\put(-45,40){\mbox{$=$}}

{\footnotesize
\put(123,17){\mbox{$ \bar S$}}
\put(17,50){\mbox{$R$}}
\put(243,22){\mbox{$\check S$}}
\qbezier(270,3)(280,20)(270,37)
\put(280,18){\mbox{$h_S = l_{S^{\vee}}=s_{_1}$}}
\qbezier(5,-5)(132,-20)(260,-5)
\put(130,-25){\mbox{$N $}}
\qbezier(5,35)(25,25)(45,35)
\put(22,20){\mbox{$l_R$}}
\qbezier(225,43)(245,52)(265,43)
\put(243,52){\mbox{$l_{\!_S}$}}
}
%{\tiny
%\qbezier(253,3)(259.5,5)(264,3)
%\put(255,8){\mbox{$l_P$}}
%}

\put(4,40){\mbox{$\ldots$}}
\put(18,40){\mbox{$\ldots$}}
\put(32,40){\mbox{$\ldots$}}

\end{picture}
%\label{compofig}
%\end{figure}

\noindent
The ordinary $N$-independent representations in this notation are $R=(R,\emptyset)$,
there conjugate are $\bar R = (\emptyset,R)$.
The simplest of non-trivial composites is the
adjoint $(1,1) = [2,1^{N-2}]$.
Vogel's universality \cite{Vogel}, providing a unified description of
representation theory of all simple
Lie algebras at once (as well as of something else)
is applicable precisely to the adjoint and its descendants (the ``$E_8$-sector"),
i.e. is one of the many topics requiring knowledge of the composites.
In knot theory, the composite representations appear in the study of
counter-strand braids, which are one of the most convenient building blocks
in the tangle calculus \cite{tangles}.

The basic special function associated with representation is its character
expressed through the Schur functions
${\rm Schur}_{(R,\emptyset)}\{{\bf p}\}={\rm Schur}_{R}\{{\bf p}\}$.
For composite representations, one needs their generalization,
composite Schur functions \cite{MMhopf}.
Because of explicit $N$ dependence, they are not easy to define for
arbitrary (generic) values of time-variable ${\bf p}_k$.
Fortunately, in most applications we need their reductions to
just $N$-dimensional loci ${\bf p}^{*V}$ (of which the simplest one
is the topological locus ${\bf p}^*_k=\frac{\{A^k\}}{\{t^k\}}$,
widely used in knot theory since \cite{DMMSS}).
At these peculiar loci, the composite Schur functions
can be defined by the uniformization trick of \cite{MMhopf},
and they possess a nice description as a bilinear combination
of the {\it skew} Schur functions
\be
{\rm Schur}_{(R,P)} = \oplus_{\eta_2=\eta_1^\vee}
{\rm Schur}_{R/{\eta_1}}\otimes {\rm Schur}_{P/\eta_2}
\label{KoikeSchur}
\ee
or, in more detail,
\be
{\rm Schur}_{(R,P)}\{{\bf p}^{*V}\} =
\sum_{\eta  } (-)^{|\eta|}\cdot {\rm Schur}_{R/\eta}\{{\bf p}^{*V}\}\cdot {\rm Schur}_{P/\eta^\vee}\{\bp^{*V}\}
\label{Koi}
\ee
This formula is due to K. Koike \cite{Koike}, see sec.6.

In the case of considering the Macdonald polynomials instead of the
Schur functions,
the uniformization still works, but it provides non-trivial expressions
with additional poles in $A$ in denominators,
e.g. already for the adjoin at the topological locus one gets
$M^*_{(1,1)} = \frac{\{Aq\}\{A\}\{A/t\}}{\{Aq/t\}\{t\}^2}$
instead of naive $\frac{\{Aq\}\{A/t\}}{\{q\}\{t\}}$,
it is this complicated expression that satisfies the uniformization
request
$\left.M^*_{(1,1)}\right|_{A=t^N} = \left.M^*_{[2,1^{N-2}]}\right|_{A=t^N}$.
Even worse, the Koike formula is not immediately deformed:
a bilinear decomposition into the skew Macdonald polynomials
survives only in the large-$A$ limit, but even then the single sum in (\ref{KoikeSchur})
restricted to $\eta_2=\eta_1^\vee$ is lifted to a double sum
with the only restriction on sizes $|\eta_2|=|\eta_1|$.

\section{Denominators}

\subsection{Example of (\ref{KerthroughDelta})
\label{exaKerthroughDelta}}

In this section, we begin from an explicit example of
what  (\ref{KerthroughDelta}) means and how it works
at a reasonably representative, but simple (and still not fully generic) level $|R|=7$.
The general claim of (\ref{KerthroughDelta}) is that
\be
\Ker^{(g)}_R\{p\} = \frac{{\rm pol}_R\{p_k,g_k\}}{\Delta_{|R|}^{(\nu_R)}\{g_k\}}
= \frac{{\rm pol}'_R\Big\{\frac{p_k}{g_k},\frac{1}{g_k}\Big\}}
{\hat\Delta_{|R|}^{(\nu_R'+1)}\Big\{\frac{1}{g_k}\Big\}}
\ \ \ \ \ \ \ \ \ \ \ \
\widehat\Ker^{(g)}_R\{p\} = \frac{\widehat{\rm pol}_R\{p_k,g_k\}}
{\hat\Delta_{|R|}^{(\hat\nu_R)}\{g_k\}}
= \frac{\widehat{\rm pol}'_R\Big\{\frac{p_k}{g_k},\frac{1}{g_k}\Big\}}
{\Delta_{|R|}^{(\hat\nu'_R+1)}\Big\{\frac{1}{g_k}\Big\}}
\label{KerthroughDelta1}
\ee
The labeling table in the case of level $7$ is
{\footnotesize
$$
\begin{array}{c|ccccccccccccccc|}
R & [1^7]&[2,1^5] & [2,2,1^3] & [2,2,2,1] & [3,1,1,1] & [3,2,1,1] & [3,2,2] & [3,3,1]
& [4,1,1,1] & [4,2,1] & [4,3] & [5,1,1] & [5,2] & [6,1] & [7]
\\
\hline
\nu_R & 1&2&3&4&5&6&7&8&9&10&11&12&13&14&15 \\
\nu'_R & 15&14&13&12&11&10&9&8&7&6&5&4&3&2&1 \\
\widehat{\nu}_R & 1&2&3&5&4&6&8&9&7&10&12&11&13&14&15\\
\widehat{\nu}'_R &15&14&13&11&12&10&8&7&9&6&4&5&3&2&1\\
\hline
\end{array}
$$
}
and the denominator functions satisfy:
{\footnotesize
$$
%\!\!\!\!\!\!\!\!\!\!\!\!\!\!\!\!\!\!
\!\!\!\!\!\!\!\!\!\!\!\!\!\!\!\!\!\!\!\!\!\!
\begin{array}{c|c|c|}
\nu'_R &&\\
\hline &&
\\
1&\Delta_{[7]}\{g\} = \Delta_{7}^{(15)}\{g\}
\sim\hat\Delta_7^{(2)}\{g^{-1}\} =\hat\Delta_{[2,1^5]}\{g^{-1}\}=\Delta_{[2,1^5]}\{g^{-1}\}
& \hat\Delta_{[7]}\{g\} = \hat\Delta_{7}^{(15)}\{g\}= \Delta_{[7]}\{g\}
\sim\Delta_7^{(2)}\{g^{-1}\} =\Delta_{[2,1^5]}\{g^{-1}\}
\\
2&\Delta_{[6,1]}\{g\} = \Delta_{7}^{(14)}\{g\}
\sim\hat\Delta_7^{(3)}\{g^{-1}\} =\hat\Delta_{[2,2,1^3]}\{g^{-1}\} =\Delta_{[2,2,1^3]}\{g^{-1}\}
& \hat\Delta_{[6,1]}\{g\} = \hat\Delta_{7}^{(14)}\{g\}= \Delta_{[6,1]}\{g\}
\sim\Delta_7^{(3)}\{g^{-1}\} =\Delta_{[2,2,1^3]}\{g^{-1}\}
 \\
&&\\
3&\Delta_{[5,2]}\{g\} = \Delta_{7}^{(13)}\{g\}
\sim\hat\Delta_7^{(4)}\{g^{-1}\} =\hat\Delta_{[3,1^4]}\{g^{-1}\}
& \hat\Delta_{[5,2]}\{g\} = \hat\Delta_{7}^{(13)}\{g\} \sim \Delta_7^{(4)}\{g^{-1}\}
=  \Delta_{[2,2,2,1]}\{g^{-1}\}
 \\
\hline
4&\Delta_{[5,1,1]}\{g\} = \Delta_{7}^{(12)}\{g\}
\sim\hat\Delta_7^{(5)}\{g^{-1}\} =\hat\Delta_{[2,2,2,1]}\{g^{-1}\}
& \hat\Delta_{[5,1,1]}\{g\} = \hat\Delta_{7}^{(11)}\{g\}
\sim   \Delta_7^{(6)}\{g^{-1}\} = \Delta_{[3,2,1,1]}\{g^{-1}\}
 \\
5&\Delta_{[4,3]}\{g\} = \Delta_{7}^{(11)}\{g\}
\sim\hat\Delta_7^{(6)}\{g^{-1}\} =\hat\Delta_{[3,2,1,1]}\{g^{-1}\}
&   \hat\Delta_{[4,3]}\{g\} = \hat\Delta_{7}^{(12)}\{g\}
\sim   \Delta_7^{(5)}\{g^{-1}\} = \Delta_{[3,1^4]}\{g^{-1}\}
\\
6&\Delta_{[4,2,1]}\{g\} = \Delta_{7}^{(10)}\{g\}
\sim\hat\Delta_7^{(7)}\{g^{-1}\} =\hat\Delta_{[4,1,1,1]}\{g^{-1}\}
&  \hat\Delta_{[4,2,1]}\{g\} = \hat\Delta_{7}^{(10)}\{g\}
\sim   \Delta_7^{(7)}\{g^{-1}\} = \Delta_{[3,2,2]}\{g^{-1}\}
\\
\hline
7&\Delta_{[4,1,1,1]}\{g\} = \Delta_{7}^{(9)}\{g\}
\sim\hat\Delta_7^{(8)}\{g^{-1}\} =\hat\Delta_{[3,2,2]}\{g^{-1}\}
&  \hat\Delta_{[4,1,1,1]}\{g\} = \hat\Delta_{7}^{(7)}\{g\}
\sim   \Delta_7^{(10)}\{g^{-1}\} = \Delta_{[4,2,1]}\{g^{-1}\}
\\
8&\Delta_{[3,3,1]}\{g\} = \Delta_{7}^{(8)}\{g\}
\sim\hat\Delta_7^{(9)}\{g^{-1}\} =\hat\Delta_{[3,3,1]}\{g^{-1}\}
&  \hat\Delta_{[3,3,1]}\{g\} = \hat\Delta_{7}^{(9)}\{g\}
\sim   \Delta_7^{(8)}\{g^{-1}\} = \Delta_{[3,3,1]}\{g^{-1}\}
\\
9&\Delta_{[3,2,2]}\{g\} = \Delta_{7}^{(7)}\{g\}
\sim\hat\Delta_7^{(10)}\{g^{-1}\} =\hat\Delta_{[4,2,1]}\{g^{-1}\}
&  \hat\Delta_{[3,2,2]}\{g\} = \hat\Delta_{7}^{(8)}\{g\}
\sim   \Delta_7^{(9)}\{g^{-1}\} = \Delta_{[4,1^3]}\{g^{-1}\}
\\
10&\Delta_{[3,2,1,1]}\{g\} = \Delta_{7}^{(6)}\{g\}
\sim\hat\Delta_7^{(11)}\{g^{-1}\} =\hat\Delta_{[5,1,1]}\{g^{-1}\}
&  \hat\Delta_{[3,2,1,1]}\{g\} = \hat\Delta_{7}^{(6)}\{g\}
\sim   \Delta_7^{(11)}\{g^{-1}\} = \Delta_{[4,3]}\{g^{-1}\}
\\
\hline
11&\Delta_{[3,1^4]}\{g\} = \Delta_{7}^{(5)}\{g\}
\sim\hat\Delta_7^{(12)}\{g^{-1}\} =\hat\Delta_{[4,3]}\{g^{-1}\}
&  \hat\Delta_{[3,1^4]}\{g\} = \hat\Delta_{7}^{(4)}\{g\}
\sim   \Delta_7^{(13)}\{g^{-1}\} = \Delta_{[5,2]}\{g^{-1}\}
\\
12&\Delta_{[2,2,2,1]}\{g\} = \Delta_{7}^{(4)}\{g\}
\sim\hat\Delta_7^{(13)}\{g^{-1}\} =\hat\Delta_{[5,2]}\{g^{-1}\}
&  \hat\Delta_{[2,2,2,1]}\{g\} = \hat\Delta_{7}^{(5)}\{g\}
\sim   \Delta_7^{(12)}\{g^{-1}\} = \Delta_{[5,1,1]}\{g^{-1}\}
\\
&& \\
13&\Delta_{[2,2,1^3]}\{g\} = \Delta_{7}^{(3)}\{g\}
\sim\hat\Delta_7^{(14)}\{g^{-1}\} =\hat\Delta_{[6,1]}\{g^{-1}\}= \Delta_{[6,1]}\{g^{-1}\}
&  \hat\Delta_{[2,2,1^3]}\{g\} = \hat\Delta_{7}^{(3)}\{g\}
\sim   \Delta_7^{(14)}\{g^{-1}\} = \Delta_{[6,1]}\{g^{-1}\} \sim \Delta_{[2,2,1^3]}\{g\}
\\
14&\Delta_{[2,1^5]}\{g\} = \Delta_{7}^{(2)}\{g\}
\sim\hat\Delta_7^{(15)}\{g^{-1}\} =\hat\Delta_{[7]}\{g^{-1}\} = \Delta_{[7]}\{g^{-1}\}
&  \hat\Delta_{[2,1^5]}\{g\} = \hat\Delta_{7}^{(2)}\{g\}
\sim   \Delta_7^{(15)}\{g^{-1}\} = \Delta_{[7]}\{g^{-1}\} \sim \Delta_{[2,1^5]}\{g\}
\\
15&\Delta_{[1^7]}\{g\} = \Delta_{7}^{(1)}\{g\} =1
&  \hat\Delta_{[ 1^7]}\{g\} = \hat\Delta_{7}^{(1)}\{g\}=1
\\
%
%&\ldots &&
\hline
\end{array}
$$
}

The proportionality signs appear because we omit monomial factors
(powers of $g_k$) appearing in the $g$-inversion of a polynomial.
Note that, in our notation, say, $\Delta_{[2,2,2,1]}=\Delta_7^{(4)}$,
but $\hat\Delta_{[2,2,2,1]}=\hat\Delta_7^{(5)}$, since the partition
$[2,2,2,1]$ has different positions in two different orderings.
As to $\hat\Delta_7^{(4)}$, it denotes $\hat\Delta_{[3,1^4]}$,
while $\Delta_{[3,1^4]}=\Delta_7^{(4)}$.
These ordering/notational details are essential for validity of (\ref{KerthroughDelta}):
it implies that
$$
\Delta_R\{g\} = \Delta^{(\nu_R)}_{|R|}\{g\}
\sim \hat\Delta^{(\nu'_R+1)}_{|R|}\{g^{-1}\}
=  \hat\Delta_{|R|}^{(\hat\nu_{R'})}\{g^{-1}\} = \hat\Delta_{R'}\{g^{-1}\}
$$
with $R'$ defined from the relation $\boxed{\hat\nu_{R'} = \nu_R'+1}$.
Likewise $R'$ is defined from
$\boxed{\nu_{R'} = \hat\nu_R'+1}$ in
$$
\hat\Delta_R\{g\} = \hat\Delta^{(\hat\nu_R)}_{|R|}\{g\}
\sim \Delta^{(\hat\nu'_R+1)}_{|R|}\{g^{-1}\}
=  \hat\Delta_{|R|}^{(\nu_{R'})}\{g^{-1}\} = \Delta_{R'}\{g^{-1}\}
$$

Actually, at level $|R|=7$ there are a few accidental (?)
coincidences, i.e. additional accidental relations between the denominator
functions.
They follow by (\ref{KerthroughDelta})
from the coincidence between the two  Kerov functions (\ref{Kerdef}) in three cases:
$R=[3,2,1,1]$, $[4,2,1]$ and $[5,2]$, when $\nu_R=\hat\nu_R$
(in addition to five generic coincidences for $[7],[6,2],[2,2,1^3],[2,1^5],[1^7]$
and to the symmetry dictated "self-duality" at $[3,3,1]$):
$$
\begin{array}{ccc}
\Delta_{[3,2,1,1]}\{g\} = \hat \Delta_{[3,2,1,1]}\{g \}
&\sim & \hat\Delta_{[5,1,1]}\{g^{-1}\}=\Delta_{[4,3]}\{g^{-1}\}
\\ \\
\hat\Delta_{[4,2,1]}\{g\}=\Delta_{[4,2,1]}\{g\}
&\sim &
\Delta_{[3,2,2]}\{g^{-1}\} = \hat\Delta_{[4,1^3]}\{g^{-1}\}
\\ \\
\hat\Delta_{[5,2]}\{g\}=\Delta_{[5,2]}\{g\}
&\sim &
\Delta_{[2,2,2,1]}\{g^{-1}\} = \hat\Delta_{[3,1^4]}\{g^{-1}\}
\end{array}
$$
and can disappear at other levels.
Only (\ref{KerthroughDelta}) is always true.

\subsection{Denominator functions}

The next addition to (\ref{Delta1to3}) helps to reveal the structure
of  denominators $\Delta_r^{(m)}$.
In the obvious abbreviated notation,
\be
\Delta^{(1)}_r=1 \nn \\ \nn \\
\Delta^{(2)}_r = \Schur_r\{g\} = S_r\nn \\ \nn \\
\Delta^{(3)}_{r}
= S_{r,r-1}+S_{r,r-2,1}-S_{r-1,r-1,1}
%= \nn \\
=  \det\left(\begin{array}{cc }
S_r & \\ & S_{r-1}
\end{array}\right)
+ S_1\cdot \det\left(\begin{array}{cc }
S_r & S_{r-1} \\  S_{r-1} & S_{r-2}
\end{array}\right)
=  \det\left(\begin{array}{ccc}
S_r & 0 & S_1 \\
S_r & S_{r-1} & 0 \\
S_{r-1}   & S_{r-2} & 1
\end{array} \right)
\nn
%\\ \nn \\
\ee
\be
\Delta^{(4)}_r
=S_2\cdot \det\left(\begin{array}{ccc}
S_r && \\ & S_{r-1} & \\ && S_{r-2}
\end{array}\right)
+S_1\cdot \det\left(\begin{array}{ccc}
S_r & S_{r-1} & S_r\cdot S_{1}^2 \\ S_{r-1} & S_{r-2} & 0 \\
S_{r-2} & S_{r-3} & S_{r-2}\cdot S_{2 }
\end{array}\right)
+ \nn \\
+S_{1}^2\cdot \det\left(\begin{array}{ccc}
S_r & S_{r-1} & 0 \\ S_{r-1} & S_{r-2} & S_{r-2}\cdot S_{2 } \\
S_{r-2} & S_{r-3} & S_{r-3}\cdot S_{1 }^2
\end{array}\right)
-S_2\cdot \det\left(\begin{array}{ccc}
S_r & S_{r-2} & 0 \\ S_{r-1} & S_{r-3} & S_{r-1}\cdot S_{ 2} \\
S_{r-2} & S_{r-4} & S_{r-2}\cdot S_{ 1 }^2
\end{array} \right)
%+ \nn \\
+ S_{2}^2S_1\cdot
\det\left(\begin{array}{ccc}
S_r & S_{r-1} & S_{r-2} \\ S_{r-1} & S_{r-2} & S_{r-3 } \\
S_{r-2} & S_{r-3} & S_{r-4}
\end{array}\right)
\nn\\
\ldots
\label{Delta1to4}
\ee

\section{Macdonald locus as an ideal in the space of $g$-polynomials
\label{MacL&R}}

We now discuss {\it the Macdonald ideal} in the ring of all polynomials of the variables $g_k$, i.e. those which vanish on the Macdonald locus. We first consider the ideal in the ring of all polynomials, and then concentrate on the ideal in the sub-ring of all homogeneous polynomials, since it is homogeneous polynomials that emerge within the Kerov polynomial context.

\subsection{Inhomogeneous polynomials}

If we parameterize the Macdonald ideal through
trigonometric functions, $g_r = \frac{\sin(ar)}{\sin(br)}$,
then all such $g_r$ can be easily expressed through $g_1$ and $g_2$.
To get these expressions, it is enough to
\begin{itemize}
\item[(i)] represent $\sin(ar)$ and $\sin(br)$
as polynomials of $\sin(a)$ and $\sin(b)$ respectively,
\item[(ii)] substitute
$\sin(a)=g_1\sin(b)$, $\cos(a) = \frac{g_2}{g_1}\cos(b)$ and, finally,
\item[(iii)] substitute $\sin^2(b) = \frac{g_2^2-g_1^2}{g_2^2-g_1^4}$.
\end{itemize}

Now one can use de Moivre's formula
\be\label{Moivre}
\sin(rx)=\sum_{k=0}^r {r!\over (r-k)!k!}\cos^kx\cdot\sin^{r-k}x\cdot\sin(r-k){\pi \over 2}
\ee
Then, for even $r=2n$,
\be
g_{2n}=g_2\cdot\frac{{\rm Pol}^{(e)}_{2n}\left(g_1^2\cdot \frac{g_2^2-g_1^2}{g_2^2-g_1^4}\right)}
{{\rm Pol}^{(e)}_{2n}\left(\frac{g_2^2-g_1^2}{g_2^2-g_1^4}\right)}
\ee
where
\be
{\rm Pol}^{(e)}_{2n}(x):=\sum_{l=0}^{n-1} {(-1)^l(2n)!\over (2(n-l)-1)!(2l+1)!}x^{n-l-1}(1-x)^l=
{(\sqrt{1-x^2}+ix)^r-(\sqrt{1-x^2}-ix)^r\over x\sqrt{1-x^2}}
\ee
Similarly, for even $r=2n+1$,
\be
g_{2n+1}=g_1\frac{{\rm Pol}^{(o)}_{2n+1}\left(g_1^2\cdot \frac{g_2^2-g_1^2}{g_2^2-g_1^4}\right)}
{{\rm Pol}^{(o)}_{2n+1}\left(\frac{g_2^2-g_1^2}{g_2^2-g_1^4}\right)}
\ee
where
\be
{\rm Pol}^{(o)}_{2n+1}(x):=\sum_{l=0}^n {(-1)^l(2n+1)!\over (2(n-l)+1)!(2l)!}x^{n-l}(1-x)^l=
{(\sqrt{1-x^2}+ix)^r-(\sqrt{1-x^2}-ix)^r\over x}
\ee
The arguments of the two polynomials in the numerator and denominator
are related by the simultaneous inversion of the two independent variables:
$g_1,g_2\longrightarrow g_1^{-1},g_2^{-1} \ \Longrightarrow \
g_1^2\cdot \frac{g_2^2-g_1^2}{g_2^2-g_1^4}\longrightarrow \frac{g_2^2-g_1^2}{g_2^2-g_1^4}$.
In result, the ideal relations are invariant under the inversion of
all $g_r \longrightarrow g_r^{-1}$.
The simplest examples are:
\be
g_3\Big((g_2^2+3g_1^4)-4g_1^2 \Big) \ \stackrel{ML}{=}\ g_1\Big(4g_1^2g_2^2-(3g_2^2+g_1^4)\Big)
\nn \\  \nn \\ \nn \\
g_4\Big((g_2^2+g_1^4)-2g_1^2 \Big)  \ \stackrel{ML}{=}\ g_2\Big(2g_1^2g_2^2- (g_2^2+ g_1^4)\Big)
\nn \\  \nn \\ \nn \\
g_5\Big((g_2^4+10g_2^2g_1^4+5g_1^8)-(12g_2^2g_1^2+20g_1^6)+16g_1^4 \Big)
 \ \stackrel{ML}{=}\
 g_1\Big(16g_2^4g_1^4 -(20g_2^4g_1^2+12g_2^2g_1^6)+(5g_2^4 + 10g_2^2g_1^4+g_1^8)\Big)
\nn \\
\ee

Now we turn to the Macdonald ideal in the sub-ring of homogeneous polynomials.
We will use the following notation:  $V_r^{(m)}$ is a homogeneous (rightly graded) polynomial of $g_1,\ldots,g_r$ of
degree $m$, which vanishes at the Macdonald locus;
${\cal V}_r^{(n)}$ is a similar polynomial, but depending only on $g_r$ and $g_1,g_2,g_3$. It is clear that, at given $r$, the minimal possible $n$ not smaller the minimal possible $m$.

\subsection{Phenomenology}

We start with a low degree examples in order to get feeling of the general structures. Let us proceed degree by degree.

\begin{itemize}

\item{
There are no functions of $g_1,g_2,g_3$ only, which vanish on ML.
}

\item{
The first vanishing combination is
\be
V_4^{(9)}:=2g_4g_3g_1^2-3g_4g_2^2g_1 +g_4g_1^5 +g_3g_2^3-3g_3g_2g_1^4+2g_2^3g_1^3
\ee
It is of degree $9$.
}

\item{
The next independent ones are at level 11
\be
V_5^{(11)} = -{4\over 3}g_1g_2g_3g_5-{1\over 2}g_2^3g_5-{4\over 3}g_1^3g_2^2g_4-{1\over 2}g_1^4g_3g_4+{5\over 6}g_2^2g_3g_4+{5\over 6}g_1^4g_2g_5+g_1^2g_4g_5+g_1^2g_2^3g_3
\ee
and
\be
V_6^{(11)}=-{1\over 6}g_1^5g_6+{1\over 6}g_1^4g_3g_4-{1\over 2}g_2^2g_3g_4+g_1^3g_2g_3^2+{1\over 3}g_1g_3^2g_4+{3\over 2}g_1g_2^2g_6-{4\over 3}g_1^2g_3g_6-g_1^2g_2^3g_3
\ee
Solving the latest 3 equations, we obtain
\be\label{sols}
g_4={g_2\over g_1}{3g_1^4g_3-2g_1^3g_2^2-g_2^2g_3\over g_1^4+2g_1g_3-3g_2^2}\\
g_5={9g_1^8g_3^2+12g_1^7g_2^2g_3-16g_1^6g_2^4-30g_1^4g_2^2g_3^2+20g_1^3g_2^4g_3+5g_2^4g_3^2
\over g_1(5g_1^8+20g_1^5g_3-30g_1^4g_2^2-16g_1^2g_3^2+12g_1g_2^2g_3+9g_2^4)}\\
g_6={g_3g_2\over g_1^2}{9g_1^4g_3-8g_1^3g_2^2-g_2^2g_3\over g_1^4+8g_1g_3-9g_2^2}
\ee}

\item{
Let us put $g_1=1$, it can be restored from grading, and let us present each $g_k$ in the form
\be
\boxed{
g_k= N_k(g_2,g_3)\cdot {P_k(g_2,g_3)\over P_k(g_2^{-1},g_3^{-1})}}
\ee
with a monomial $N_k(g_2,g_3)$ of $g_2$, $g_3$ and a polynomial $P_k(g_2,g_3)$:
\be
\begin{array}{rlrl}
N_4&=-{1\over g_2g_3},& P_4&=g_2^2g_3+2g_2^2-3g_3\\
&&&\\
N_5&={1\over g_2^4g_3^2},& P_5&=5g_2^4g_3^2+20g_2^4g_3-16g_2^4-30g_2^2g_3^2+12g_2^2g_3+9g_3^2\\
&&&\\
N_6&=-{1\over g_2},& P_6&=g_2^2g_3+8g_2^2-9g_3
\end{array}
\label{homrat}
\ee
so that
\be
{\cal V}_k \sim g_1^{-k} g_k N_k^{-1}(g_1^2g_2^{-1},g_1^3g_3^{-1})
P_k(g_1^2g_2^{-1},g_1^3g_3^{-1}) -  P_k(g_1^{-2}g_2,g_1^{-3}g_3)
\ee
}

\item{The next non-trivial level is 14, where $g_7$ and $g_8$ first emerge.

Their manifest form is given by
\be\label{sols2}
\begin{array}{rlrl}
N_7&=-{1\over g_2^6g_3^3},& P_7&=7g_2^6g_3^3+ 84g_2^6g_3^2-105g_2^4g_3^3-64g_2^6-168g_2^4g_3^2+192g_2^4g_3+189g_2^2g_3^3-108g_2^2g_3^2-27g_3^3\\
&&&\\
N_8&=-{1\over g_2^5g_3^3},& P_8&= P_4\cdot (g_2^4g_3^2+16g_2^4g_3-8g_2^4-18g_2^2g_3^2+9g_3^2)
\end{array}
\ee
}

\item{
At level 11, there is a set of vanishing combinations that involve $V_4^{(9)}$, $V_5^{(11)}$ and $V_6^{(11)}$, they are linear combinations of 4 basic elements, say, of
\be
g_1^2V_4^{(9)};\ \ \ \ \ V_5^{(11)};\ \ \ \ \ V_6^{(11)};\ \ \ \ \ g_2V_4^{(9)}
\ee
At level 14, the structure of vanishing combinations is much similar: there are 15 basic elements which can be parted to
\be
\hspace{-1.5cm}\hbox{\bf A:}\ \ \ \ \ \{g_5,g_1g_4,g_1^2g_3,g_2g_3,g_1g_2^2,g_1^3g_2,g_1^5\}V_4^{(9)};\ \ \ \ \ \{g_3,g_1g_2,g_1^3\}V_5^{(11)};
\ \ \ \ \ \{g_3,g_1g_2,g_1^3\}V_6^{(11)}
\ee
{\bf B:} any two relations giving $g_7$ and $g_8$, e.g.,
\be
\hspace{-1.3cm}
\begin{array}{rl}
V_7^{(14)}&={2\over 5}g_1^6g_2g_6+{5\over 3}g_1^4g_4g_6-{4\over 3}g_1g_3^2g_7+{12\over 5}g_1^2g_5g_7+g_2^2g_3g_7-{7\over 3}g_1^4g_3g_7+{7\over 2}g_1^3g_2^2g_7+g_1^3g_3^2g_5-{2\over 3}g_1^5g_2^2g_5%+\nn\\
+{1\over 2}g_1^6g_3g_5-\\
&\\
&-{7\over 30}g_1^7g_7-3g_1^3g_2g_3g_6-2g_1^3g_2g_4g_5-{7\over 5}g_1g_2g_5g_6+{1\over 6}g_1^2g_2^2g_3g_5-3g_1g_2g_4g_7+{7\over 3}g_1g_3g_4g_6+g_1^2g_2g_3^2g_4\\
&\\
V_8^{(14)}&=54g_1^4g_2^2g_3^2-48g_1^3g_2^4g_3-3g_2^4g_3^2-8g_1g_2^3g_3g_4+21g_1^2g_2g_3^2g_4-9g_2^3g_8
-8g_1^6g_4^2+2g_3^2g_4^2-16g_1^4g_2^3g_4
-24g_1^3g_3g_4^2+\\
&\\
&+12g_1^2g_2^2g_4^2+63g_1^2g_2^3g_6+9g_2^2g_4g_6+3g_1^4g_2g_8+18g_1^2g_4g_8
+20g_1^4g_4g_6-84g_1^3g_2g_3g_6%-\nn\\
-12g_1g_2g_3g_8-8g_1g_3g_4g_6
\end{array}
\ee
}

\item{
This scheme further changes: $g_{10}$ emerge at level 16, $g_9$, at level $17$, $g_{12}$ at level 18, and $g_{11}$ at level 20. Hence, the odd $g_{2k+1}$ first appears at level $3k+5$, while the even ones, $g_{2k}$, less regularly: $k=2,3,4,5,6,7$ emerge at levels $9,11,14,16,18,21$ accordingly so that

{\footnotesize
%\qquad{-1cm}
\be
\hspace{-1cm}
\begin{array}{rlrl}
N_9&=-{1\over g_2^6g_3^2},&
P_9&=g_2^6g_3^3+24g_2^6g_3^2+48g_2^6g_3-27g_2^4g_3^3-64g_2^6-144g_2^4g_3^2
+144g_2^4g_3+99g_2^2g_3^3-72g_2^2g_3^2-9g_3^3\\
&&&\\
N_{10}&={1\over g_2^7g_3^4},& P_{10}&=(g_2^4g_3^2+28g_2^4g_3+16g_2^4-30g_2^2g_3^2-60g_2^2g_3+45g_3^2)
(5g_2^4g_3^2+20g_2^4g_3-16g_2^4-30g_2^2g_3^2+12g_2^2g_3+9g_3^2)\\
&&&\\
N_{11}&=-{1\over g_2^{10}g_3^5},&
P_{11}&=11g_2^{10}g_3^5+440g_2^{10}g_3^4+2288g_2^{10}g_3^3-495g_2^8g_3^5-704g_2^{10}g_3^2-6336g_2^8g_3^4
-2816g_2^{10}g_3-4752g_2^8g_3^3+\\
&&&\\
&&&+4158g_2^6g_3^5+1024g_2^{10}+12672g_2^8g_3^2+14256g_2^6g_3^4-2304g_2^8g_3-14256g_2^6g_3^3-8910g_2^4g_3^5
-1728g_2^6g_3^2+\\
&&&\\
&&&+6480g_2^4g_3^3+4455g_2^2g_3^5-3240g_2^2g_3^4-243g_3^5\\
&&&\\
N_{12}&={1\over g_2^7g_3^3},&
P_{12}&=P_4\cdot P_6\cdot (g_2^4g_3^2+40g_2^4g_3-32g_2^4-42g_2^2g_3^2+24g_2^2g_3+9g_3^2)
\end{array}
\label{NP}
\ee
}}

\item{
The solution polynomials $P_K$ can be compactly rewritten through the two functions
\be
K_1:=6(g_3-g_2^2);\ \ \ \ \ \ K_2:=P_4=g_2^2g_3+2g_2^2-3g_3
\ee
in the following way:
\be\label{PK}
P_4&=&K_2\nn\\
P_5&=&5K_2^2-K_1^2\nn\\
P_6&=&K_2-K_1\nn\\
P_7&=&7K_2^3-K_1^3-7K_1K_2^2-7K_1^2K_2\nn\\
P_8&=&P_4\cdot (K_2^2-K_1^2-2K_1K_2)\nn\\
P_9&=&K_2^3+{1\over 3}K_1^3-3K_1K_2^2-K_1^2K_2\\
P_{10}&=&P_5\cdot (K_2^2-K_1^2-4K_1K_2)\nn\\
P_{11}&=&11K_2^5+K_1^5-55K_2^4K_1-22K_2^3K_1^2+22K_2^2K_1^3+11K_1^4K_2\nn\\
P_{12}&=&P_4\cdot P_6\cdot (K_2^2-3K_1^2-6K_1K_2)\nn
\ee
These polynomials depend on the choice of the normalization factor $N_k$. We shall see in the next subsection what is the natural choice of $N_k$ and how to get general formulas for the polynomials $P_k$. In particular, it turns out that, in the case of $P_{2n}$, it is better to make another choice of $N_{2n}$, while the choice made here for $N_{2n+1}$ is good. This is why the polynomials $P_{2n+1}$ have a clear structure here.
}

\end{itemize}

From these formulas, one can immediately read off:

{\footnotesize
\be
%\!\!\!\!\!\!\!\!\!\!\!\!\!\!\!\!\!\!\!\!\!
\!\!\!\!\!\!\!\!\!\!\!
\begin{array}{cl}
{\cal V}_4^{(9)} =&
g_1^5g_4  -3g_1^4g_2g_3+2g_1^3g_2^3+2g_1^2g_3g_4  -3g_1g_2^2g_4  +g_2^3g_3
\nn \\ \nn \\
{\cal V}_5^{(14)} =&
 5g_1^9g_5  -9g_1^8g_3^2-12g_1^7g_2^2g_3+16g_1^6g_2^4+20g_1^6g_3g_5  -30g_1^5g_2^2g_5
 +30g_1^4g_2^2g_3^2-20g_1^3g_2^4g_3-16g_1^3g_3^2g_5  +12g_1^2g_2^2g_3g_5  +9g_1g_2^4g_5  -5g_2^4g_3^2  \nn \\ \nn \\
{\cal V}_6^{(12)} =&
g_1^6g_6  -9g_1^4g_2g_3^2+8g_1^3g_2^3g_3+8g_1^3g_3g_6  -9g_1^2g_2^2g_6  +g_2^3g_3^2
\nn\\ \nn \\
{\cal V}_7^{(21)} =&
 7g_1^{14}g_7  -27g_1^{12}g_3^3-108g_1^{11}g_2^2g_3^2+192g_1^{10}g_2^4g_3-64g_1^9g_2^6
 +84g_1^{11}g_3g_7  -105g_1^{10}g_2^2g_7  +189g_1^8g_2^2g_3^3
 -\nn \\ &
 -168g_1^7g_2^4g_3^2-168g_1^7g_2^2g_3g_7  +189g_1^6g_2^4g_7  -105g_1^4g_2^4g_3^3+84g_1^3g_2^6g_3^2
 -64g_1^5g_3^3g_7  +192g_1^4g_2^2g_3^2g_7  -108g_1^3g_2^4g_3g_7  -27g_1^2g_2^6g_7  +7g_2^6g_3^3
 \nn \\ \nn \\
{\cal V}_8^{(23)} =&
g_1^{15}g_8  -27g_1^{12}g_2g_3^3+18g_1^{11}g_2^3g_3^2+24g_1^{10}g_2^5g_3-16g_1^9g_2^7
+18g_1^{12}g_3g_8  -21g_1^{11}g_2^2g_8  +63g_1^8g_2^3g_3^3
-84g_1^7g_2^5g_3^2+24g_1^6g_2^7g_3+24g_1^9g_3^2g_8
-\nn \\ &
-84g_1^8g_2^2g_3g_8  +63g_1^7g_2^4g_8
-21g_1^4g_2^5g_3^3+18g_1^3g_2^7g_3^2-16g_1^6g_3^3g_8  +24g_1^5g_2^2g_3^2g_8  +18g_1^4g_2^4g_3g_8
-27g_1^3g_2^6g_8  +g_2^7g_3^3 \nn \\ \nn \\
{\cal V}_9^{(24)} =&
g_1^{15}g_9  -9g_1^{12}g_3^4-72g_1^{11}g_2^2g_3^3+144g_1^{10}g_2^4g_3^2-64g_1^9g_2^6g_3
+24g_1^{12}g_3g_9  -27g_1^{11}g_2^2g_9  +99g_1^8g_2^2g_3^4
-144g_1^7g_2^4g_3^3
+48g_1^6g_2^6g_3^2+48g_1^9g_3^2g_9
-\nn \\ &
-144g_1^8g_2^2g_3g_9  +99g_1^7g_2^4g_9  -27g_1^4g_2^4g_3^4
+24g_1^3g_2^6g_3^3-64g_1^6g_3^3g_9  +144g_1^5g_2^2g_3^2g_9
-72g_1^4g_2^4g_3g_9
-9g_1^3g_2^6g_9  +g_2^6g_3^4
\nn \\ \nn \\ \nn \\
\ldots
\end{array}
\ee
}

\subsection{Systematic approach: gauge $g_1=1$}

One can parameterize the Macdonald locus through $g_r = \frac{\sin(ar)}{\sin(br)}
\cdot{\sin^r b\over\sin^r a}$ so that $g_1=1$. Then, one can easily get that
\be
\left.\sin^2a={3(g_3-g_2^2)\over g_2^2g_3-4g_2^2+3g_3}={K_1\over 2(K_1+K_2)},\ \ \ \ \ \ \ \ \sin^2b=\sin^2a\right|_{g_{2,3}\to g_{2,3}^{-1}}
\ee
Now one can use de Moivre's formula (\ref{Moivre}) in order to obtain formulas (\ref{PK}), for instance,
\be
\sin(2n+1)a=\sin(a)\cdot\sum_{k=0}^n {(-1)^{n-k}(2n+1)!\over (2n+1-2k)!(2k)!}\cos^{2k}a\cdot\sin^{2n-2k}a=\nn\\
=\sin a\cdot\sum_{k=0}^n {(-1)^{n-k}(2n+1)!\over (2n+1-2k)!(2k)!}
\Big(1-{K_1\over 2(K_1+K_2)}\Big)^k\cdot\left({K_1\over 2(K_1+K_2)}\right)^{n-k}=\nn\\
=\sin a\cdot \Big(2(K_1+K_2)\Big)^{-n}\cdot\sum_{k=0}^n {(-1)^{n-k}(2n+1)!\over (2n+1-2k)!(2k)!}
\Big(2K_2+K_1\Big)^k\cdot K_1^{n-k}
\ee
and
\be
g_{2n+1}={\sin(2n+1)a\over\sin(2n+1)b}{\sin^{2n+1}b\over\sin^{2n+1}a}=\Big({\bar K_1\over K_1}\Big)^n\cdot
{\sum_{k=0}^n {(-1)^k(2n+1)!\over (2n+1-2k)!(2k)!}
\Big(2K_2-K_1\Big)^k\cdot K_1^{n-k}\over \sum_{k=0}^n {(-1)^k(2n+1)!\over (2n+1-2k)!(2k)!}
\Big(2\bar K_2-\bar K_1\Big)^k\cdot \bar K_1^{n-k}}=\nn\\
={(-1)^n\over g_2^{2n}g_3^n}\cdot{\sum_{k=0}^n {(-1)^k(2n+1)!\over (2n+1-2k)!(2k)!}
\Big(2K_2-K_1\Big)^k\cdot K_1^{n-k}\over \sum_{k=0}^n {(-1)^k(2n+1)!\over (2n+1-2k)!(2k)!}
\Big(2\bar K_2-\bar K_1\Big)^k\cdot \bar K_1^{n-k}}
\ee
where $\bar K_{1,2}:=K_{1,2}\Big|_{g_{2,3}\to g_{2,3}^{-1}}$. Thus, in this case,
\be\label{odd}
\boxed{
N_{2n+1}={(-1)^n\over g_2^{2n}g_3^n},\ \ \ \ \ \ \ P_{2n+1}=\sum_{k=0}^n {(-1)^k(2n+1)!\over (2n+1-2k)!(2k)!}
\Big(2K_2-K_1\Big)^k\cdot K_1^{n-k}}
\ee
These formulas agree with (\ref{PK}) up to inessential factor of $2^{-n}$ for an exception of $P_9$, which becomes
\be
P_9=9K_2^4+K_1^4-18K_1^2K_2^2-24K_1K_2^3
\ee
upon choosing the normalization factor $N_9={1\over g_2^8g_3^4}$ instead of that in formula (\ref{NP}), where the normalization factor was chosen $N_9=-{1\over g_2^6g_3^2}$.

Similarly,
\be
\sin(2na)=\sin a\cdot\cos a\cdot\sum_{k=1}^n {(-1)^{n-k}(2n)!\over (2n-2k+1)!(2k-1)!}\cos^{2(k-1)}a\cdot\sin^{2(n-k)}a=\nn\\
=\sin a\cdot\cos a\cdot \Big(2(K_1+K_2)\Big)^{1-n}\cdot\sum_{k=1}^n {(-1)^{n-k}(2n)!\over (2n-2k+1)!(2k-1)!}
\Big(2K_2+K_1\Big)^{k-1}\cdot K_1^{n-k}
\ee
and
\be
g_{2n}={\sin(2na)\over\sin(2nb)}{\sin^{2n}b\over\sin^{2n}a}=(-1)^{n+1}g_2^{3-2n}g_3^{1-n}\cdot{\sum_{k=1}^n {(-1)^{k}(2n)!\over (2n-2k+1)!(2k-1)!}
\Big(2K_2+K_1\Big)^{k-1}\cdot K_1^{n-k}\over\sum_{k=1}^n {(-1)^{k}(2n)!\over (2n-2k+1)!(2k-1)!}
\Big(2\bar K_2+\bar K_1\Big)^{k-1}\cdot \bar K_1^{n-k}}
\ee
where we used that
\be
{\sin a\cos a\over\sin b\cos b}=-g_3g_2^3\cdot{\bar K_1+\bar K_2\over K_1+K_2}
\ee
Thus, we can see that, in this case, it is better to choose the normalization factor $N_{2n}=(-1)^{n+1}g_2^{3-2n}g_3^{1-n}$:
\be\label{even}
\boxed{
N_{2n}=(-1)^{n+1}g_2^{3-2n}g_3^{1-n},\ \ \ \ \ \ \ P_{2n}=\sum_{k=1}^n {(-1)^{k}(2n)!\over (2n-2k+1)!(2k-1)!}
\Big(2K_2+K_1\Big)^{k-1}\cdot K_1^{n-k}}
\ee
Note that $N_6$ and $N_{12}$ in (\ref{NP}) are different, and, hence, the corresponding $P_6$ and $P_{12}$ in (\ref{PK}) differs from (\ref{even}).

\section{Composite Kerov and Macdonald functions
}

As we discussed in subsection \ref{Comp},
straightforward is the definition of Kerov/Macdonald/Schur functions
for $N$-dependent conjugate and composite representations
on the Miwa locus $p^X$ for $N$ Miwa variables.
The full-fledged functions $\widetilde\Ker_{([R],[P])}\{{\bf p}^{*V}_k(A,g_k),\bp^{*V}_k(A,g_k),g_k\}$
in the composite representation $([R],[P])$ are then introduced by an
"uniformization" procedure {\it a la} \cite{univ},
as a kind of an analytical continuation:
\be
\Ker_{([R],[P])}^{(g)}[X]\to \widetilde{\Ker}_{([R],[P])}\{{\bf p}^{*V}_k(A,g_k),\bp^{*V}_k(A,g_k),g_k\}
\label{uniker}
\ee
with the functions ${\bf p}^{*V}_k(A,g_k)$, $\bp^{*V}_k(A,g_k)$ yet to be defined.
According to this definition, the uniform Kerov function may explicitly
depend on $N$, and, indeed, it is a non-trivial and even non-polynomial
function of $A=t^N$.

In this section, we go through particular examples on increasing complexity
with the goal to illustrate the structure of at least the l.h.s. of
(\ref{uniker}), already this being a non-trivial task.

\subsection{Conjugate representations $\bar S = (\emptyset,S)$}

The simplest under conjugation of the Young diagram
is the behaviour of the Schur functions at the Miwa locus:
what is transformed is the locus itself,
\be
\Schur_{\bar S} [X] = {\cal X}^{l_{S^\vee}}\cdot \Schur_{ S} [X^{-1}]
\label{Schurconj}
\ee
where ${\cal X}:=\det X = \prod_{a=1}^n x_a$.

Already in this example it is clear that a uniform
$\Ker_{\bar S}^{(g)}\{p_k\}$ will not be easy to define,
because $ X^{-1}$ has no clear relation to $p_k$ on the Miwa locus
(traces are not consistent with inversion).

The ``$U(1)$-factor" ${\cal X}^{l_{S^\vee}}$ in (\ref{Schurconj})  can be eliminated by
restriction from $GL(N)$ to $SL(N)$, i.e. by further restricting
the Miwa locus to ${\cal X}=\det X=1$.
It is, however, useful to keep this factor, because it sheds
additional light on the structure of the Kerov deformation.
The power of ${\cal X}$ is defined by the sum of sizes of
$S$ and $\bar S$: the former is $|S|$ but the latter depends on $N$
and the number $l_{S^\vee}$ of lines in transposed $S$, as clear
from the picture in sec.\ref{Comp}:
$|\bar S| = N\cdot l_{S^\vee}-|S|$.
Taking the sum instead of the difference is explained by inversion
of $X$ at the r.h.s. of (\ref{Schurconj}).

Already the deformation of (\ref{Schurconj}) is non-trivial:
the r.h.s. contains several terms
all with the same power
of ${\cal X}$,
\be
\Ker_{\bar S}^{(g)}[X] =  {\cal X}^{l_{S^\vee} } \cdot
\sum_{Q\vdash |S| }
  B_{SQ}^{(g)}\cdot\Ker_{Q}^{(g)}[X^{-1}]
\label{conjuKerexpansion}
\ee
the sum runs over the diagrams $Q$ of the size $|S|$ (what is denoted by $\vdash$).
This structure will be further inherited by formulas for generic composite
representations:
after the Kerov deformation, the
Koike formula (\ref{KoikeSchur}) acquires new terms as compared to the Schur case.

In the case of antisymmetric representations $S=[1^s]$ with $S^\vee=[s]$ and $l_{S^\vee}=1$,
there is just a single new term at the r.h.s., and (\ref{Schurconj}) remains un-deformed:
\be
\boxed{
\Ker_{[1^s]}^{(g)}[X] =  {\cal X}  \cdot \Ker_{[1^{N-s}]}^{(g)}[X^{-1}]
}
\ee
However, things change already for the symmetric representations $S=[s]$.
Before going deeper into this story in sec.\ref{symrep} and further,
we consider a couple of simple composite examples,
where no {\it new} structure constants emerge as compared to the Schur case.

In what follows we denote $\bar X:=X^{-1}$ to simplify the formulas.
If appears, $g^{-1}$ means inversion of all $g_k\longrightarrow g_k^{-1}$.

\subsection{Adjoint representation $adj=([1],[1])$}

Adjoint is the simplest of composite representations,
it is described by the Young diagram $[2,1^{N-2}]$.
In this simplest case,
\be
\boxed{
\Ker_{([1],[1])}^{(g)}[X] = {\cal X} \cdot \left(\Ker^{(g)}_{[1]}[X]\cdot \Ker^{(g)}_{[1]}[\bar X]
- \frac{\Schur_{[1]}\{g\}\cdot \Schur_{[N-1]}\{g\}}{\Schur_{[N]}\{g\}}\right)
}
\label{ker(1,1)}
\ee
Denominator here comes from $\Ker_{([1],[1])}^{(g)}=\Ker_{[2,1^{N-2}]}^{(g)}$, which is the
second ($\nu_{[2,1^{N-2}]}=2$) Kerov function of the weight $N$, and has $\Delta^{(2)}_N\{g\}=
{\Schur}_N\{g\}$ as a denominator.

(\ref{ker(1,1)}) can be easily promoted to
\be
\Ker_{([1],[1])}^{(g)}\{{\bf p}^{*x}_k,\bp^{*x}_k  |N\} = \Ker^{(g)}_{[1]}\{{\bf p}^{*x}_k\}\Ker^{(g)}_{[1]}\{\bp^{*x}_k\}
- \frac{\Schur_{[1]}\{g\}\cdot \Schur_{[N-1]}\{g\}}{\Schur_{[N]}\{g\}}
= {\bf p}^{*x}_1\bp^{*x}_1 - \frac{\Schur_{[1]}\{g\}\cdot \Schur_{[N-1]}\{g\}}{\Schur_{[N]}\{g\}}
\ee
but a somewhat non-trivial $N$-dependence emerges, which is not easy to
express through the uniform parameter $A$.

However, this can be easily done on the Macdonald locus, where
\be
\frac{\Schur_{[1]}^{\Mac}\cdot \Schur_{[N-1]}^{\Mac}}{\Schur_{[N]}^{\Mac}}
= \frac{ \{q\} \cdot \{t^{N}\}}{\{t\}\{qt^{N-1}\}}
= \frac{\{q\}\{A\}}{\{t\}\{Aq/t\}}
\ee
But even this simple expression is {\bf non-polynomial in $A$}.
In result, on the intersection of topological and Macdonald loci (TML), one gets
\be
\Ker^{**}_{([1],[1])}(A,q,t) = \Mac^*_{\rm adj}(A,q,t) =
\left(\frac{\{A\}}{\{t\}}\right)^2 - \frac{\{q\}\{A\}}{\{t\}\{Aq/t\}}
= \frac{\{Aq\}\{A\}\{A/t\}}{\{Aq/t\} \{t\}^2}
= \frac{\{A\}\{q\}}{\{Aq/t\}\{t\}}\cdot \frac{\{Aq\}\{A/t\}}{\{q\}\{t\}}
\ee
rather than just $\frac{\{Aq\}\{A/t\}}{\{q\}\{t\}}$, which one could
naively (but erroneously) expect.

\subsection{The series $([1],[1^s])$}

\be
\boxed{
\Ker_{([1],[1^s])}^{(g)}[X] =
{\cal X}\cdot\left(
\Ker^{(g)}_{[1]}[X]\cdot\Ker^{(g)}_{[1^s]}[\bar X]
- \frac{\Schur_{[1]}\{g\}\cdot \Schur_{[N-s]}\{g\}}{\Schur_{[N+1-s]}\{g\}}
\cdot \Ker_{[1^{s-1}]}^{(g)}[\bar X]
\right)
}
\ee
In this case $(1,[1^s]) = [2,1^{N-s-1}]$, which is again the second Kerov function
($\nu_{[2,1^{N-s-1}]}=2$), but this time, of weight $N+1-s$, thus the denominator
is $\Delta^{(2)}_{N+1-s}\{g\} = \Schur_{[N+1-s]}\{g\}$.

\subsection{Conjugation of symmetric representation
\label{symrep}}

We new return to the conjugate representations.
As already mentioned, a new structure constant emerges already for the
simplest symmetric representation $[2]$:
\be
\boxed{
\Ker_{(\emptyset,[2])}^{(g)}[X] = \Ker_{\overline{[2]}}^{(g)}[X] =
{\cal X}^2\cdot\left(
\Ker^{(g)}_{[2]}[\bar X]
+ B_{[2],[1,1]}^{(N)}\!\{g\} \cdot \Ker^{(g)}_{[1,1]}[\bar X]
\right)
}
\ee
where $B_{[2],[1,1]}^{(N=2)}\{g\}=0$, but already at $N=3$
\be
B_{[2],[1,1]}^{(3)}\!\{g\} = 2\cdot
\frac{2g_4g_3g_1^2-3g_4g_2^2g_1 +g_4g_1^5 +g_3g_2^3-3g_3g_2g_1^4+2g_2^3g_1^3}
{\Schur_{[2]}\{g\}\cdot \Delta^{(3)}_4\{g\}}
= \frac{2V_4^{(9)}\{g\}}
{\Schur_{[2]}\{g\}\cdot \Delta^{(3)}_4\{g\}}
\ee
The numerator
$V_4^{(9)}\{g\}$
in this expression is an element of the Macdonald ideal, as we explained in the previous section.
For $N=4$
\be
B_{[2],[1,1]}^{(4)}\!\{g\} = \frac{2V_6^{(19) }\{g\}}{ \Schur_{[2]}\{g\}\cdot \Delta^{(4)}_6\{g\}}
\ee
with an even more sophisticated polynomial in the numerator, which is again an element of the Macdonald ideal, etc.
For general $N$
\be
\boxed{
B_{[2],[1,1]}^{(N)}\!\{g\} = \frac{2V_{2N-2}\{g\}}{ \Schur_{[2]}\{g\}\cdot \Delta^{(N)}_{2N-2}\{g\}}
}
\ee
They all have zero grading.

The denominator  in this formula is a product of those for $\Ker_{(\emptyset,[2])}^{(g)}$
and $\Ker_{ [2] }^{(g)}$.
The latter one is just $\Schur_{[2]}\{g\}$, and the former one is defined from the fact
that $(\emptyset,[2])=[2^{N-1}]$ is the $N$-th Kerov function of weight $2N-2$,
thus it is equal to $\Delta^{(N)}_{2N-2}\{g\}$.

\subsection{Conjugation of higher symmetric representation
\label{symreps}}

At $N=2$ for all $s$ one has just $B_{[s],Q}^{(N=2)} = \delta_{Q,[s]}$, i.e.
\be
{\rm for \ \ }N=2: \ \ \ \ \ \  \Ker_{(\emptyset,[s])}^{(g)}[X] = \Ker_{\overline{[s]}}^{(g)}[X] =
{\cal X}^s \cdot \Ker^{(g)}_{[2]}[\bar X]
\label{conjatN2}
\ee
exactly like the Schur case.
Note that, in the case of Kerov functions, these symmetric $[s]$ do {\it not} exhaust all
independent representations at $N=2$: as we discussed in sec.\ref{Miwa}, some of the Kerov functions labeled by Young diagrams with 3 lines are non-zero.

However, at all other $N$ the deviation from the Schur case is quite significant.
At $N=3$
\be
\Ker_{(\emptyset,[3])}^{(g)}[X] = \Ker_{\overline{[3]}}^{(g)}[X] =
{\cal X}^3\cdot\left(
\Ker^{(g)}_{[3]}[\bar X]
+ B_{[3],[2,1]}^{(N)}\!\{g\} \cdot \Ker^{(g)}_{[2,1]}[\bar X]
+  B_{[3],[1,1,1]}^{(N)}\!\{g\} \cdot \Ker^{(g)}_{[1,1,1]}[\bar X]
\right)
\ee
and already the denominators are slightly different in the two coefficients:
\be
B_{[3],[2,1]}^{(3)} = \frac{3V_6^{(35)}\{g\}}{\underbrace{6g_1^3g_2g_3 \cdot\Schur_{[3]}\{g^{-1}\}}_{\Delta_3^{(3)}\{g\}} \cdot \Delta^{(7)}_6\{g\}},
\ \ \ \ \ \ \
B_{[3],[1,1,1]}^{(3)} =
\frac{3V_6^{(33)}\{g\}}{12\cdot\Schur_{[3]}\{g\} \cdot \Delta^{(7)}_6\{g\}}
\ee
with
\be
\Delta^{(7)}_6\{g\} \sim \Delta^{(6)}_6\{g^{-1}\}
\label{Delta33inB}
\ee
of grading $30$, which coincides with the denominator of $\Ker_{[3,3]}$.

Note that one can solve, say, the condition $B_{[2],[1,1]}^{(3)}=0$ in order to determine $g_4$ as a function of $(g_1,g_2,g_3)$, then solve the two similar conditions $B_{[3],[1,2]}^{(3)}=0$ and $B_{[3],[1,1,1]}^{(3)}=0$ to determine $g_5,g_6$ as functions of $(g_1,g_2,g_3)$, etc. Thus, one obtains all higher $g_k$'s as functions of three arbitrary parameters $(g_1,g_2,g_3)$. It turns out that solving these conditions, one unambiguously led to the Macdonald polynomials with the parameters $q$ and $t$ obtained from the equations
\be
g_2=g^{\Mac}_2\left({g_1\over g^{\Mac}_1}\right)^2,\ \ \ \ \ \ g_3=g^{\Mac}_3\left({g_1\over g^{\Mac}_1}\right)^3
\ee
with $g^{\Mac}_k = \frac{\{q^k\}}{\{t^k\}}$. All other $g_k$ are then
\be
g_k=g^{\Mac}_k\left({g_1\over g^{\Mac}_1}\right)^k
\ee
The parameter $g_1$ remains unfixed, since the transformation of measure $g_k\to \xi^kg_k$ with arbitrary $\xi$ does not change the symmetric polynomials. Thus, the requirement of absence additional structure constants is equivalent to the Macdonald ideal.

For generic $N$ and $s$, the denominator of $B^{(N)}_{[s],R}$ is a product $\Schur_{[s]}\{g^{-1}\}\cdot \Delta^{(\nu_{[s^{N-1}]})}_{N(s-1)}\{g\}
\sim \Schur_{[s]}\{g^{-1}\}\cdot \Delta^{(\nu'_{[s^{N-1}]}+1)}_{N(s-1)}\{g^{-1}\}$.
In (\ref{Delta33inB}), we have $\nu_{[3,3]}=7$ and $\nu'_{[3,3]}+1=5+1=6$.
In fact, there can be partial cancellations with the denominator of
$\Ker_{R}\{g\}$, which goes into the numerator of $B^{(N)}_{[s],R}$,
but they are not very essential.

%\newpage???
\subsection{Adjoint tower}

To already considered
\be
\Ker_{([1],[1])}^{(g)}[X] = {\cal X} \cdot \left(\Ker^{(g)}_{[1]}[X]\cdot \Ker^{(g)}_{[1]}[\bar X]
- a_{[1],\emptyset}\right)
\ee
with
\be
a_{[1],\emptyset} = \frac{\Schur_{[1]}\{g\}\cdot \Schur_{[N-1]}\{g\}}{\Schur_{[N]}\{g\}}
\ee
we can add
\be
N=2: &
\Ker_{([2],[2])}^{(g)}[X] = {\cal X}^2 \cdot \left(\Ker^{(g)}_{[2]}[X]\cdot \Ker^{(g)}_{[2]}[\bar X]
-\alpha_{[2],[1]} \cdot\Ker^{(g)}_{[1]}[X]\cdot \Ker^{(g)}_{[1]}[\bar X]
+ \alpha_{[2],\emptyset} \right)
\ee
with
\be
\alpha_{[2],[1]} ={4\over g_1^2g_3g_4}\cdot\frac{g_4g_3+3g_4g_2g_1+3g_3g_2g_1^2-g_4g_1^3}
{\Schur_{[2]}\{g\}\cdot\Schur_{[4]}\{g^{-1}\}}= {4\over g_1^2g_3g_4}\cdot\frac{g_4g_3+3g_4g_2g_1+3g_3g_2g_1^2-g_4g_1^3}
{\Schur_{[2]}\{g\}\cdot\Delta_{4}^{(\nu_{[4]} )}\{g \}}\nn \\
\alpha_{[2],\emptyset}  = \frac{2g_4g_3-8g_4g_1^3+3g_3g_2^2+6g_3g_2g_1^2-3g_3g_1^4}
{g_3g_4\cdot\Schur_{[2]}\{g\}^2\cdot\Schur_{[4]}\{g^{-1}\}}=
\frac{2g_4g_3-8g_4g_1^3+3g_3g_2^2+6g_3g_2g_1^2-3g_3g_1^4}
{g_3g_4\cdot\Schur_{[2]}\{g\}^2\cdot\Delta_{4}^{(\nu_{[4]} )}\{g \}}
\ee
This $\alpha_{[2],\emptyset}  $ vanishes at the {\it Schur} locus (but not in the Macdonald ideal), where all $g_r=1$,
but $\alpha_{[2],[1]}  $ does not.

For $N>2$, there are more terms at the r.h.s:
\be\label{2}
\Ker_{([2],[2])}^{(g)}[X] = {\cal X}^2 \cdot \left(\Ker^{(g)}_{[2]}[X]\cdot \Ker^{(g)}_{[2]}[\bar X]
+ a_{[2],[1,1]}^{(2|N)}\cdot\Ker^{(g)}_{[2]}[X]\cdot \Ker^{(g)}_{[1,1]}[\bar X]
+ a_{[1,1],[2]}^{(2|N)}\cdot\Ker^{(g)}_{[1,1]}[X]\cdot \Ker^{(g)}_{[2]}[\bar X]
+ \right. \nn \\   \left.
+ a_{[1,1],[1,1]}^{(2|N)}\cdot\Ker^{(g)}_{[1,1]}[X]\cdot \Ker^{(g)}_{[1,1]}[\bar X]
+a_{[1],[1]}^{(2|N)}\cdot\Ker^{(g)}_{[1]}[X]\cdot \Ker^{(g)}_{[1]}[\bar X]
+ a_{\emptyset,\emptyset}^{(2|N)}\right)
\ee
The denominators of $a_{R,R'}^{(2|N)}$
are proportional to
$\Delta_{2N}^{(\nu'_{[4,2^{N-2}]}+1 )}\{g^{-1}\}
\sim \Delta_{2N}^{(\nu_{[4,2^{N-2}]} )}\{g \} $ and to the square
of ${\rm Schur}_{[2]}\{g^{-1}\}$ coming from the product of two $\Ker_{[2]}$, from the first term at the r.h.s. in (\ref{2}).
At $N=3$, $a_{[1,1],[1,1]}^{(2|3)}$ is undistinguishable from $a_{[1],[1]}^{(2|3)}$.
Also $a_{[2],[1,1]}^{(2|3)}\sim V_6^{(15)}$ and $a_{[1,1],[2]}^{(2|3)}\sim V_6^{(19)}$
belong to the Macdonald ideal.

Generally,
\be
&\Ker_{([s],[s])}^{(g)}[X] = {\cal X}^s \cdot\sum_{r\leq s} \left(
\sum_{R',R''\vdash r } a_{R',R''}^{(s|N)}\cdot
\Ker^{(g)}_{[R']}[X]\cdot \Ker^{(g)}_{[R'']}[\bar X]
 \right)&
 \nn \\
&\Biggr\downarrow \ \ \lefteqn{\rm Macdonald\ Locus}& \nn \\ \nn \\
&\Mac_{([s],[s])} [X] = {\cal X}^s \cdot\sum_{r\leq s} \left(
  {\cal B}_{r}^{(s)}(A,q,t)\cdot
\Mac_{[r]}[X]\cdot \Mac_{[r]}[\bar X]
 \right)&
 \nn \\
 &\Biggr\downarrow \ \ \lefteqn{\rm Schur\ Locus}& \nn \\ \nn \\
&{\rm Schur}_{([s],[s])} [X] = {\cal X}^s \cdot  \left(
{\rm Schur}_{[s]}[X]\cdot {\rm Schur}_{[s]}[\bar X]
- {\rm Schur}_{[s-1]}[X]\cdot {\rm Schur}_{[s-1]}[\bar X]
 \right)&
 \nn \\
\ee
Moreover, {\bf the uniformization occurs at ML}, but it is not simple to express the coefficients $a_{r}^{(s|N)}$
in this formula through the ratios of products of Schur functions:
since, at the ML, the latter look like ratios of products of $\{qt^i\}$ or $\{q^jt\}$ with various $i$ and $j$, and
$a_{r}^{(s|N)}$ involve the factors like $\{q^it^j\}$ at concrete $N$, $A=t^N$, as we shall see in the next section.

\section{Uniformization at Macdonald locus}

In this section, we demonstrate what happens to Kerov functions
in composite representations at the Macdonald locus,
how the uniformization emerges
and what explains further simplifications to the Koike formula
after restriction to the Schur locus $t=q$.

For the Schur functions, i.e. for $t=q$, for the arbitrary composite representation
 $(R,P)$ made from a pair of Young diagrams $R$ and $P$ and
for $A=q^N$ and $V$ with no more than $N$ elements
(diagrams $S$ with no more than $N$ lines), there is the Koike formula (\ref{Koi})
\be
{\rm Schur}_{(R,P)}\{{\bf p}^{*V}\} =
\sum_{\eta  } (-)^{|\eta|}\cdot {\rm Schur}_{R/\eta}\{{\bf p}^{*V}\}\cdot {\rm Schur}_{P/\eta^\vee}\{\bp^{*V}\}
\label{compoSchur}
\ee
where ${\bf p}^{*V}$, $\bp^{*V}$ are calculated at $t=q$, the sum goes over Young sub-diagrams $\eta$.
Note that $\eta$ in the second factor is transposed so that only $\eta\subset R\cap P^\vee$ contribute.

In the case of Macdonald functions, the situation gets more involved (see also \cite{CheEl}).
There are three basic modifications of (\ref{compoSchur}):

{\bf (i)} The skew Macdonald polynomials instead of the skew Schur functions are expected to be sufficient only
in the limit of $N\longrightarrow \infty$, which is interpreted  as $A=t^N\longrightarrow 0$
at $|t|<1$. This limit coincides with the limit of $A\longrightarrow\infty$.

{\bf (ii)} The sum turns into a double sum over arbitrary diagrams $\eta_1$ and $\eta_2$
of equal sizes, but without the requirement $\eta_2=\eta_1^\vee$.

{\bf (iii)} In the sum, there emerge non-unit coefficients that are functions of $q$ and $t$.
Those in front of the items with $\eta_2\neq \eta_1^\vee$ are suppressed by the
factor $\{q/t\}$ (in fact, it is a more interesting factor measuring the
distance between $\eta_2$ and $\eta_1^\vee$).

In other words, (\ref{compoSchur}) is substituted by
\be
M_{(R,P)} \{{\bf p}^{*V}\} =
\sum_{\stackrel{\zeta_1,\zeta_2}{|R|-|\zeta_1|=|P|-|\zeta_2|}  }
(-)^{|\eta|}{\cal B}^{\zeta_1,\zeta_2}_{(R,P)}(A,q,t)
\cdot M_{\zeta_1}\{{\bf p}^{*V}\}\cdot
M_{\zeta_2}\{\bp^{*V}\}
\approx \nn \\
\approx
\sum_{\stackrel{\eta_1,\eta_2}{|\eta_1|=|\eta_2|}  }
\left(-\frac{t}{q}\right)^{|\eta|}\!\!\!\cdot B^{\eta_1,\eta_2}_{(R,P)}(q,t)
\cdot M_{R/\eta_1}\{{\bf p}^{*V}\}\cdot
M_{P/\eta_2}\{\bp^{*V}\} + O(A^{2})
\label{compoMac}
\ee
Note that the expansion parameter $-\frac{q}{t} = {\bf t}$ is exactly the
same as in the Poincare polynomials of the Khovanov-Rozansky complexes
used in the definition of superpolynomials \cite{DGR}.

In the remaining part of this section we provide examples
of (\ref{compoMac}) for various cases, however, a general formula for the coefficients ${\cal B}^{\zeta_1,\zeta_2}_{(R,P)}(A,q,t)$
is still missed.

\subsection{Conjugate representations}

It turns out that the property of conjugation representations (\ref{Schurconj}) is correct not only on Schur,
but, for a wide class of representations $\mathfrak{S}$, also on the whole Macdonald locus in the space ${\cal G}$ of the time variables $g_k$, and
\be\label{conj}
\Mac_{S} [X] =  {\cal X}^{l_{S^\vee}} \cdot \Mac_{\bar S} [X^{-1}]\ \ \ \ \ \ \ S\in \mathfrak{S}
\ee
It remains to describe $\mathfrak{S}$. It turns out that {\bf $\mathfrak{S}$ is the set of Young diagrams that consist of no more then two rectangles.}

Let us note that, at any concrete $N$, the Young diagram $(R,P)$ is conjugate to $(P,R)$ at the same $N$. For instance, $([2,1],[2])$ at $N=4$ is the Young diagram $[4,3,2]$, and its conjugate $\overline{[4,3,2]}=[4,2,1]$ is just $([2],[2,1]$ at $N=4$. In this case, the property (\ref{conj}) is not satisfied. It also means that, when (\ref{conj}) is satisfied, i.e. for rectangular $R$ and $P$, there is an identity
\be
\Mac_{(R,P)} [X] =  {\cal X}^{l_{S^\vee}} \cdot \Mac_{(P,R)} [X^{-1}]\ \ \ \ \ \ \ R\hbox{ and }P\hbox{ are rectangular}
\ee
For instance, it is correct for $R$ and $P$ being symmetric and antisymmetric representations.
It implies that the corresponding formulas for symmetric and antisymmetric composite Macdonald polynomials in the next subsection turns to be related with each other.

\subsection{Answers for symmetric and antisymmetric representations}

For symmetric and antisymmetric representations, there are general formulas for the coefficients ${\cal B}^{\zeta_1,\zeta_2}_{(R,P)}(A,q,t)$.

\subsubsection*{\framebox{Representation $([r],[p]) = [r+p,p^{N-2}]$}}

\be
%\boxed{
M_{([r],[p])}\{{\bf p}^{*S}\} =
\sum_{i=0}^{{\rm min}(r,p)} {\cal B}_{([r],[p])}^{(i)}\cdot M_{[r-i]}\{{\bf p}^{*S}\}
\cdot M_{[p-i]}\{\bp^{*S}\}
%}
\ee
with
\be
\boxed{
{\cal B}_{([r],[p])}^{(i)} =  \frac{\{Aq^{r+p-2i} \}}{ \{ Aq^{r+p}   \}} \cdot \prod_{j=1}^i \left(\frac{\Big\{\frac{q^{j-1}}{t}\Big\}}{\{q^j\}}\cdot
\frac{\{q^{r-j+1}\}\{q^{p-j+1}\}}{\{q^{r-j}t\}\{q^{p-j}t\}}\cdot
\frac{\{Aq^{r+p-j+1} \}}{\Big\{\frac{Aq^{r+p-j} }{t}\Big\}}\right)
} \nn \\
\approx \left(\frac{t}{q}\right)^i\cdot \prod_{j=1}^i \left(\frac{\Big\{\frac{q^{j-1}}{t}\Big\}}{\{q^j\}}\cdot
\frac{\{q^{r-j+1}\}\{q^{p-j+1}\}}{\{q^{r-j}t\}\{q^{p-j}t\}}\right) + O(A^{2})
\label{Brp}
\ee
The ratio in front of the product in this formula substitutes the multiplier $\{ Aq^{r+p}\}$ in the product  by
$\{ Aq^{r+p-2i}\}$.

This formula is an illustration of {\bf (ii)} and {\bf (iii)}:
the Schur level selection rule $\eta_2=\eta_1^\vee$ is violated,
but deviations are damped by peculiar factorials $\Big\{\frac{q^{i-1}}{t}\Big\}!$,
not just by $\Big\{\frac{q}{t}\Big\}$ as one could expect.

\subsubsection*{\framebox{Representation $([r],[1^p])=[r+1, 1^{N-p-1}]$}}

\be
%\boxed{
M_{([r],[1^p])}\{{\bf p}^{*S}\} =
\sum_{i=0}^{{\rm min}(r,p)} {\cal B}_{([r],[1^p])}^{(i)}\cdot M_{[r-i]}\{{\bf p}^{*S}\}
\cdot M_{[1^{p-i}]}\{\bp^{*S}\}
%}
\ee
with
\be
\boxed{
{\cal B}_{([r],[1^p])}^{(i)} =
(-)^i\frac{\Big\{\frac{Aq^{r-i}}{t^{p-i}}\Big\}}{\Big\{\frac{Aq^{r}}{t^{p}}\Big\}}
\prod_{j=r+1-i}^r \frac{\{q^j\}}{\{q^{j-1}t\}}
\ \approx \ \left(-\frac{t}{q}\right)^i\cdot \prod_{j=p+1-i}^p \frac{\{q^j\}}{\{q^{j-1}t\}}
}
\label{Br1p}
\ee

\subsubsection*{\framebox{Representation $([1^r],[p])=[(p+1)^r,p^{N-r-1}]$}}

\be
%\boxed{
M_{([1^r],[ p])}\{{\bf p}^{*S}\} =
\sum_{i=0}^{{\rm min}(r,p)} {\cal B}_{([1^r],[ p])}^{(i)}\cdot M_{[1^{r-i}]}\{{\bf p}^{*S}\}
\cdot M_{[{p-i}]}\{\bp^{*S}\}
%}
\ee
with
\be
\boxed{
{\cal B}_{([1^r],[p])}^{(i)} = B_{([p],[1^r])}^{(i)} \ \stackrel{(\ref{Br1p})}{=} \
(-)^i\frac{\Big\{\frac{Aq^{p-i}}{t^{r-i}}\Big\}}{\Big\{\frac{Aq^{p}}{t^{r}}\Big\}}
\prod_{j=p+1-i}^p \frac{\{q^j\}}{\{q^{j-1}t\}}
\ \approx \ \left(-\frac{t}{q}\right)^i\cdot \prod_{j=p+1-i}^p \frac{\{q^j\}}{\{q^{j-1}t\}}
}
\label{B1rp}
\ee

\subsubsection*{\framebox{Representation $([1^r],[1^p])=[2^{r},1^{N-r-p}]$}}

\be
%\boxed{
M_{([1^r],[ 1^p])}\{{\bf p}^{*S}\} =
\sum_{i=0}^{{\rm min}(r,p)} {\cal B}_{([1^r],[ 1^p])}^{(i)}\cdot M_{[1^{r-i}]}\{{\bf p}^{*S}\}
\cdot M_{[1^{p-i}]}\{\bp^{*S}\}
%}
\ee
with
\be
\boxed{
{\cal B}_{([1^r],[1^p])}^{(i)} = (-)^i\cdot
\frac{\Big\{\frac{A}{t^{r+p-2i}}\Big\}}{\Big\{\frac{A}{t^{r+p-i}}\Big\}}
\prod_{j=1}^i \left(\frac{\Big\{\frac{q}{t^{j-1}}\Big\}}{\{t^j\}}\cdot
\frac{\Big\{\frac{A}{t^{r+p-j}}\Big\}}{\Big\{\frac{Aq}{t^{r+p-j}}\Big\}}\right)
\ \approx \ \left(-\frac{t}{q}\right)^i\cdot
\prod_{j=1}^i \left(\frac{\Big\{\frac{q}{t^{j-1}}\Big\}}{\{t^j\}}\right)
}
\ee
In variance with (\ref{Brp}), this expression
does not automatically vanish for $i>{\rm min}(r,p)$.
The role of the ratio in front of the product is similar to that in (\ref{Brp}).

\subsection{The case of $R=[2,1]$}

When $R=[2,1]$, the composite Macdonald polynomials already can not be presented by a combination of the skew Macdonald polynomials. This is a new phenomenon, and we discuss it in some details.

\subsubsection*{\framebox{Representation $([2,1],[p]) = [p+2,p+1,p^{N-3}]$}}

In this case, the general expression is
\be
M_{([2,1],[p])}\{{\bf p}^{*S}\} =   M_{[2,1]}\{{\bf p}^{*S}\}\cdot M_{[p]}\{\bp^{*S}\}
-  \Big({\cal B}_{([2,1],[p])}^{[2]} \cdot M_{[2]}\{{\bf p}^{*S}\} + {\cal B}_{([2,1],[p])}^{[1,1]}\cdot M_{[1,1]}\{{\bf p}^{*S}\}\Big)
\cdot M_{[p-1]}\{\bp^{*S}\} + \nn \\
+ {\cal B}_{([2,1],[p])}^{[1]}\cdot M_{[1]}\{{\bf p}^{*S}\} \cdot M_{[p-2]}\{\bp^{*S}\}
- {\cal B}_{([2,1],[p])}^{\emptyset}\cdot M_{[p-3]}\{\bp^{*S}\}
\ee
and the coefficients can be calculated:
\be
{\cal B}_{([2,1],[p])}^{[2]}= \frac{\{q^p\}}{\{q^{p-1}t\}}\cdot
\frac{\Big\{\frac{Aq^{p-1}}{t}\Big\}}{\Big\{\frac{Aq^p}{t^2}\Big\}}, \ \ \ \ \ \ \
{\cal B}_{([2,1],[p])}^{[1,1]} = \frac{\{q^p\}\{t^2\}\{q^2t\}}{\{q^{p-1}t\}\{qt\}\{qt^2\}}\cdot
\frac{\{Aq^p\}}{\Big\{\frac{Aq^{p+1}}{t}\Big\}} \nn \\
{\cal B}_{([2,1],[p])}^{[1]} = {\{q^p\}\{q^{p-1}\}\over \{q^{p-1}t\}\{q^{p-2}t\}}\cdot
{\{t^2\}\{q^3\}\{qt\}\{Aq^p\}\{{Aq^{p-1}\over t}\}
- \{{q\over t}\}\{q\}\{q^2t^2\} \{{Aq^{p+1}\over t}\}\{Aq^{p-2}\}
\over\{q^2\}\{qt\}\{qt^2\}{Aq^p\over t^2}\}\{{Aq^{p+1}\over t}\}}
\nn\\
{\cal B}_{([2,1],[p])}^{\emptyset}= {\{q^{p-2}\}\{q^{p-1}\}\{q^p\}\{t/q\}\{t^2\}\over\{q^{p-3}t\}\{q^{p-2}t\}\{q^{p-1}t\}\{qt^2\} \{t\}}
\cdot{\{Aq^{p+1}\}\{Aq^{p-3}\}\over\{{Aq^{p+1}\over t}\}\{{Aq^p\over t^2}\}}
\ee
One can see from these formulas that the combination
\be
{\cal B}_{([2,1],[p])}^{[2]}M_{[2]} + {\cal B}_{([2,1],[p])}^{[1,1]}M_{[1,1]}
= \frac{\{q^p\}}{\{q^{p-1}t\}}\cdot
\left(\frac{\Big\{\frac{Aq^{p-1}}{t}\Big\}}{\Big\{\frac{Aq^p}{t^2}\Big\}} \cdot M_{[2]}
+ \frac{\{Aq^p\}}{\Big\{\frac{Aq^{p+1}}{t}\Big\}}\cdot
\frac{\{t^2\}\{q^2t\}}{\{qt\}\{qt^2\}}\cdot M_{[1,1]}\right)
\ee
is not proportional to $M_{[2,1]/[1]}$ for finite $A$. However, in the limit of small (or large) $A$, this skew Macdonald polynomial emerges:
\be
{\cal B}_{([2,1],[p])}^{[2]}M_{[2]} + {\cal B}_{([2,1],[p])}^{[1,1]}M_{[1,1]}
\approx \frac{\{q^p\}}{\{q^{p-1}t\}}\cdot \frac{t}{q}\cdot M_{[2,1]/[1]} + O(A^{2})
\ee
Let us stress that the coefficient ${\cal B}_{([2,1],[p])}^{[1]}$ in front of $M_{[1]}$ is not factorized, which is not that surprising,
because even in the small $A$ limit, when it is expressed through the skew Macdonald polynomials, it is a combination of $M_{[2,1]/[2]}$ and $M_{[2,1]/[1,1]}$. It can be also equivalently presented in the form
\be
{\cal B}_{([2,1],[p])}^{[1]} = { \{q^p\}\{q^{p-1}\}\over \{q^{p-1}t\}\{q^{p-2}t\}}\cdot
\left(
{ \{t^2\}\{q^3\}\{qt\} \over \{q^2\}\{qt\}\{qt^2\} } \cdot
{\{Aq^p\}\{ {Aq^{p-1}\over t} \} \over \{ {Aq^p\over t^2} \}\{ {Aq^{p+1}\over t}\} }
- {\{ {q\over t} \}\{q\}\{q^2t^2\}\over\{q^2\}\{qt\}\{qt^2\}}\cdot
{ \{Aq^{p-2}\}\over \{{Aq^p\over t^2}\} } \right)
\ee

%\newpage???

\subsubsection*{\framebox{Representation $([2,1],[1^p]) = [3,2,1^{N-p-2}]$}
\label{21asp}}

\be
M_{([2,1],[1^p])}\{{\bf p}^{*S}\} =   M_{[2,1]}\{{\bf p}^{*S}\}\cdot M_{[1^p]}\{\bp^{*S}\}
-  \Big({\cal B}_{([2,1],[1^p])}^{[2]} \cdot M_{[2]}\{{\bf p}^{*S}\}
+ {\cal B}_{([2,1],[1^p])}^{[1,1]}\cdot M_{[1,1]}\{{\bf p}^{*S}\}\Big)
\cdot M_{[1^{p-1}]}\{\bp^{*S}\} + \nn \\
+ {\cal B}_{([2,1],[1^p])}^{[1]}\cdot M_{[1]}\{{\bf p}^{*S}\} \cdot M_{[1^{p-2}]}\{\bp^{*S}\}
- {\cal B}_{([2,1],[1^p])}^{\emptyset}\cdot M_{[1^{p-3}]}\{\bp^{*S}\}
\label{21aspdeco}
\ee
with
\be\label{109}
{\cal B}_{([2,1],[1^p])}^{[2]} = \frac{\{q\}}{\{t\}}\cdot
\frac{\Big\{\frac{A}{t^s}\Big\}}{\Big\{\frac{Aq}{t^{s+1}}\Big\}}
& \approx \ \frac{t}{q}\cdot\frac{\{q\}}{\{t\}} = \frac{1-q^{-2}}{1-t^{-2}} \nn \\
{\cal B}_{([2,1],[1^p])}^{[1,1]} = \frac{\{q\}}{\{t\}}\cdot \frac{\{t^2\}\{q^2t\}}{\{qt\}\{qt^2\}}
\frac{\Big\{\frac{Aq}{t^{s-1}}\Big\}}{\Big\{\frac{Aq^2}{t^{s }}\Big\}}
& \approx \ \frac{t}{q}\cdot\frac{\{q\}}{\{t\}} \cdot \frac{\{t^2\}\{q^2t\}}{\{qt\}\{qt^2\}} = \frac{1-q^{-2}}{1-t^{-2}}\cdot \frac{\{t^2\}\{q^2t\}}{\{qt\}\{qt^2\}} \nn \\
{\cal B}_{([2,1],[1^p])}^\emptyset =
\Big\{\frac{q}{t}\Big\}\cdot \frac{\{q\}\{q^2\}}{\{t\}^2\{qt^2\}} \cdot
\frac{\Big\{\frac{A}{t^{s-3}}\Big\}\Big\{\frac{A}{t^{s+1}}\Big\}}
{\Big\{\frac{Aq^2}{t^{s}}\Big\}\Big\{\frac{Aq}{t^{s+1}}\Big\}}
& \approx \ \left(\frac{t}{q}\right)^3 \cdot
\Big\{\frac{q}{t}\Big\}\cdot \frac{\{q\}\{q^2\}}{\{t\}^2\{qt^2\}}
\ee
$$
{\cal B}_{([2,1],[1^p])}^{[1]} = \frac{\{q\}}{\{t\}}\cdot\frac{1}{\{qt^2\}}\cdot
\frac{1}{\Big\{\frac{Aq}{t^{s+1}}\Big\}\Big\{\frac{Aq^2}{t^s}\Big\}}\cdot\left(
\left(\{q^2t\}+\frac{\{q^2\}\{q/t\}}{\{q\}}\right)\cdot \Big\{\frac{A}{t^s}\Big\}\Big\{\frac{Aq}{t^{s-1}}\Big\}
+ \frac{\{t\}\{q/t\}^2\{q^2t^2\}}{\{qt\}}
\right)
$$
Clearly, the combination
\be
{\cal B}_{([2,1],[1^p])}^{[2]} M_{[2]} + {\cal B}_{([2,1],[1^p])}^{[1,1]} M_{[1,1]}
\ \approx \ \frac{t}{q}\cdot\frac{\{q\}}{\{t\}}\cdot M_{[2,1]/[1]} + O(A^2)
\ee
but it is not proportional to the skew Macdonald polynomial at finite $A$.
Moreover, the deviation depends on $s$, i.e. is {\it not universal}.
Note that at this level, there is just {\it one} skew Macdonald polynomial,
thus one can not cure the deviation from it by taking linear combinations.

In particular, at $q=t$ the last item in (\ref{21aspdeco})
does not contribute: $\alpha^\emptyset$ vanishes.

Also at $q=t$ the penultimate term should be
${\rm Schur}_{[2,1]/[2]}\cdot {\rm Schur}_{[1^s]/[1,1]}$,
while ${\rm Schur}_{[2,1]/[1,1]}$ does {\it not} contribute.
Then, since
\be
M_{[2,1]/[2]} = p_1= M_{[1]}, \ \ \ \ \ \ \ \
M_{[2,1]/[1,1]} =\frac{\{t^2\}\{q^2t\}}{\{qt\}\{qt^2\}}\cdot p_1
= \frac{\{t^2\}\{q^2t\}}{\{qt\}\{qt^2\}}\cdot M_{[1]}
\ee
it implies the decomposition
\be
{\cal B}_{([2,1],[1^p])}^{[1]} \approx 1 \oplus \Big\{\frac{q}{t}\Big\}\cdot \frac{\{t^2\}\{q^2t\}}{\{qt\}\{qt^2\}}
\ee
Decomposition with such a property indeed exists, but not unique.
Say, one can convert ${\cal B}_{([2,1],[1^p])}^{[1]}$ in (\ref{109}) into
\be
{\cal B}_{([2,1],[1^p])}^{[1]} = \frac{\{q\}}{\{t\}} \cdot
\frac{1}{\Big\{\frac{Aq}{t^{s+1}}\Big\}\Big\{\frac{Aq^2}{t^s}\Big\}}\cdot\left(
\left(\frac{\{q^2\}}{\{qt\}}+\frac{ \{q/t\}\{q^2t\}}{\{qt\}\{qt^2\}}\right)\cdot \Big\{\frac{A}{t^s}\Big\}\Big\{\frac{Aq}{t^{s-1}}\Big\}
+ \frac{\{t\}\{q/t\}^2\{q^2t^2\}}{\{qt\}\{qt^2\}}
\right)
\ee
or into
\be
{\cal B}_{([2,1],[1^p])}^{[1]} = \frac{\{t^3\}\{q^2\}\{q\}}{\{qt^2\}\{t^2\}\{t\}}\cdot
\frac{\Big\{\frac{A}{t^s}\Big\}\Big\{\frac{Aq}{t^{s-1}}\Big\}}
{\Big\{\frac{Aq}{t^{s+1}}\Big\}\Big\{\frac{Aq^2}{t^{s}}\Big\}}
+ \frac{\Big\{\frac{q}{t}\Big\}\{q^2t^2\}\{q\}}{\{t^2\}\{qt\}\{qt^2\}}\cdot
\frac{ \Big\{\frac{A}{t^{s-2}}\Big\}}{ \Big\{\frac{Aq^2}{t^{s}}\Big\}}
\ee
or into other similar expressions.

\subsubsection*{\framebox{Representation $([2,1],[2,1]) = [4,3,2^{N-4},1]$}}

\be
M_{([2,1],[2,1])}\{{\bf p}^{*S}\} =
M_{[2,1]}\{{\bf p}^{*S}\}\cdot M_{[2,1]}\{\bp^{*S}\}
- {\cal B}_{([2,1],[2,1])}^{[2]}\cdot M_{[2]}\{{\bf p}^{*S}\}\cdot M_{[2]}\{\bp^{*S}\}
- {\cal B}_{([2,1],[2,1])}^{[1,1]}\cdot M_{[1,1]}\{{\bf p}^{*S}\}\cdot M_{[1,1]}\{\bp^{*S}\}
- \!\!\!\!\!\!\!\!\!\! \nn \\
\!\!\!\!\!\!\!\!\!\!
-{\cal B}_{([2,1],[2,1])}^{[2],[1,1]}\cdot\Big(M_{[2]}\{{\bf p}^{*S}\}\cdot M_{[1,1]}\{\bp^{*S}\}
+ M_{[1,1]}\{{\bf p}^{*S}\}\cdot M_{[2]}\{\bp^{*S}\}\Big)
%+ \nn \\
+ {\cal B}_{([2,1],[2,1])}^{[1]}\cdot M_{[1]}\{{\bf p}^{*S}\}\cdot M_{[1]}\{\bp^{*S}\}
- {\cal B}_{([2,1],[2,1])}^\emptyset \ \ \ \ \ \ \
\ee
with
\be
{\cal B}_{([2,1],[2,1])}^{[2]} =\frac{\{q\}}{\{t\}}\cdot\frac{\Big\{\frac{A}{t^2}\Big\}}{\Big\{\frac{Aq}{t^3}\Big\}}
\ \approx \ \frac{t}{q}\frac{\{q\}}{\{t\}}
\nn \\
{\cal B}_{([2,1],[2,1])}^{[1,1]}=
\frac{\{q\} }{\{t\} }\cdot
\left(\frac{\{t^2\}\{q^2t\}}{\{qt\}\{qt^2\}}\right)^2
\cdot\frac{\{Aq^2\}}{\Big\{\frac{Aq^3}{t}\Big\}}
\ \approx \ \frac{t}{q}\frac{\{q\}}{\{t\}} \cdot
\left(\frac{\{t^2\}\{q^2t\}}{\{qt\}\{qt^2\}}\right)^2
\nn \\ \nn \\
{\cal B}_{([2,1],[2,1])}^{[2],[1,1]} =
\frac{\{q\} }{\{t\} }\cdot
\frac{\{t^2\}\{q^2t\}}{\{qt\}\{qt^2\}}
\cdot\frac{\Big\{\frac{Aq}{t}\Big\}}{\Big\{\frac{Aq^2}{t^2}\Big\}}
\ \approx \ \frac{t}{q}\frac{\{q\}}{\{t\}}\cdot
\frac{\{t^2\}\{q^2t\}}{\{qt\}\{qt^2\}}
\nn 
\\
{\cal B}_{([2,1],[2,1])}^{[1]}=
{\{q\}\Big\{\frac{Aq^3}{t}\Big\}\Big\{\frac{Aq^2}{t^2}\Big\}^2\Big\{\frac{Aq}{t^3}\Big\}\over\{t\}}
\Biggr(
{\{q^2t\}\over t^2q\}}\Big\{\frac{Aq^2}{t^2}\Big\}\Big\{\frac{Aq}{t}\Big\}\Big(\{Aq^2\}\Big\{\frac{Aq}{t^3}\Big\}+
\Big\{\frac{Aq^3}{t}\Big\}\Big\{\frac{A}{t^2}\Big\}\Big)-\nn\\
-{\Big\{\frac{q}{t}\Big\}^2\over \{t^2q\}^2}
\Biggr[\Big({\{q^2\}\{t^2\}\over \{q\}\{t\}}
+{\{t^2q^2\}\over\{tq\}}\Big)\Big({q^6A^4\over t^6}+{t^6\over q^6A^4}\Big)-{\{q^2t^2\}^2\over\{qt\}^2}
\Big((q^4t^4+q^2t^4+q^4+q^2t^2+t^2+1)\cdot{qA^2\over t^5}+\nn\\
+(q^4t^4+q^4t^2+t^4+q^2t^2+q^2+1)\cdot{t\over q^5A^2}\Big)+
{1\over q^5t^5}\Big(q^{10}t^{10}+q^{10}t^8+q^8t^{10}+3q^8t^6+3q^6t^8+q^8t^4+q^6t^6+q^4t^8+\nn\\
\left.+6q^6t^4+6q^4t^6+q^6t^2+
q^4t^4+q^2t^6+3q^4t^2+3q^2t^4+q^2+t^2+1\Big)\Biggr]\right)
\nn
\ee
\be\label{1212}
{\cal B}_{([2,1],[2,1])}^{\emptyset} = \frac{\{q\}^2}{\{t\}^2}\cdot
\frac{\{A\}}{\Big\{\frac{Aq}{t^3}\Big\}\Big\{\frac{Aq^2}{t^2}\Big\}^2\Big\{\frac{Aq^3}{t}\Big\}}
\left({\{q^2t\}\over \{t^2q\}} \Big\{\frac{Aq}{t}\Big\}\Big\{\frac{Aq}{t^3}\Big\}
\Big\{\frac{Aq^3}{t}\Big\} - {\Big\{{q\over t}\Big\}^2\{q^2\}\{t^2\}\over\{q\}\{t\}\{t^2q\}^2}
\Big\{{Aq^2\over t}\Big\}\Big\{{Aq^2\over t^2}\Big\}\Big\{{Aq\over t^2}\Big\}+\right.\nn\\
\left.+{\Big\{{q\over t}\Big\}^2\{q^2t\}\over \{t^2q\}}\Big(\{q^2t^2\}
\Big\{{Aq^2\over t^2}\Big\}+\Big\{{Aq\over t}\Big\}\Big)
\right)\ \approx \ 
{t^3\over q^3}\frac{\{q\}^2}{\{t\}^2}\cdot
\left({\{q^2t\}\over \{t^2q\}} - {\Big\{{q\over t}\Big\}^2\{q^2\}\{t^2\}\over\{q\}\{t\}\{t^2q\}^2}
\right)
\nn\\
\ee
It follows that
\be
{\cal B}_{([2,1],[2,1])}^{[2]}\cdot M_{[2]}\{{\bf p}^{*S}\}\cdot M_{[2]}\{\bp^{*S}\}
+{\cal B}_{([2,1],[2,1])}^{[2],[1,1]}\cdot\Big(M_{[2]}\{{\bf p}^{*S}\}\cdot M_{[1,1]}\{\bp^{*S}\}
+ M_{[1,1]}\{{\bf p}^{*S}\}\cdot M_{[2]}\{\bp^{*S}\}\Big)+\nn\\
+ {\cal B}_{([2,1],[2,1])}^{[1,1]}\cdot M_{[1,1]}\{{\bf p}^{*S}\}\cdot M_{[1,1]}\{\bp^{*S}\}
\approx \
\frac{t}{q}\frac{\{q\}}{\{t\}} \cdot M_{[2,1]/[1]}\{{\bf p}^{*S}\}\cdot M_{[2,1]/[1]}\{\bp^{*S}\}
+ O(A^{2})\nn
\ee
as expected.

Note that, in this example, we first meet a new property: ${\cal B}_{([2,1],[2,1])}^{\emptyset}$ is not factorized even in the small $A$ limit despite it is proportional to the only skew Macdonald term $M_{[2,1]/[2,1]}=1$. Indeed, the first and the second terms in ${\cal B}_{([2,1],[2,1])}^{\emptyset}$ survive in this limit. At the same time, the Schur case is clearly reproduced in (\ref{1212}): upon specialization $t=q$, the second and the third terms in ${\cal B}_{([2,1],[2,1])}^{\emptyset}$ vanish, and one immediately gets ${\cal B}_{([2,1],[2,1])}^{\emptyset}=1$.

\subsection{The case of $[r,1]$}

This is a generalization of the  subsection \ref{21asp} from $r=2$ to arbitrary $r$.

\subsubsection*{\framebox{Representation $([r,1],[1^p]) = [r+1,2,1^{N-p-2}]$}}

\be
M_{([r,1],[1^p])}\{{\bf p}^{*S}\} =   M_{[r,1]}\{{\bf p}^{*S}\}\cdot M_{[1^p]}\{\bp^{*S}\}
- \nn \\
+ \sum_{m=1}^{r-1} (-)^m \Big({\cal B}_{([r,1],[1^p])}^{[r+1-m]} \cdot M_{[r+1-m]}\{{\bf p}^{*S}\}
+ {\cal B}_{([r,1],[1^p])}^{[r-m,1]}\cdot M_{[r-m,1]}\{{\bf p}^{*S}\}\Big)
\cdot M_{[1^{p-m}]}\{\bp^{*S}\}  + \nn \\
+(-)^r {\cal B}_{([r,1],[1^p])}^{[1]} \cdot M_{[1]}\{{\bf p}^{*S}\}\cdot M_{[1^{p-r}]}\{\bp^{*S}\}
\ +\  (-)^{r+1}  {\cal B}_{([r,1],[1^p])}^{\emptyset}\cdot M_{[1^{p-r-1}]}\{\bp^{*S}\}
\ee
with
\be
{\cal B}_{([r,1],[1^p])}^{[r]} = \frac{\{q\}}{\{t\}}\cdot
\frac{\Big\{\frac{A}{t^s}\Big\}}{\Big\{\frac{Aq}{t^{s+1}}\Big\}}
& \approx \ \frac{t}{q}\cdot\frac{\{q\}}{\{t\}} = \frac{1-q^{-2}}{1-t^{-2}}
\nn \\
{\cal B}_{([r,1],[1^p])}^{[r-m,1]} = \frac{ \{q^rt\}\{q^{r-m-1}t^2\}}
{ \{q^{r-m}t\} \{q^{r-1}t^2\}}
\left(\prod_{j=1}^m \frac{\{q^{r-j}\}}{\{q^{r- j-1}t\}}\right)
\cdot \frac{\Big\{\frac{Aq^{r-m}}{t^{s-m}}\Big\}}{\Big\{\frac{Aq^{r}}{t^{s }}\Big\}}
& \approx \
\left(\frac{t}{q}\right)^m \cdot
\frac{ \{q^rt\}\{q^{r-m-1}t^2\}}
{ \{q^{r-m}t\} \{q^{r-1}t^2\}}
\left(\prod_{j=1}^m \frac{\{q^{r-j}\}}{\{q^{r- j-1}t\}}\right)
\nn \\
{\cal B}_{([r,1],[1^p])}^\emptyset =
\frac{\Big\{\frac{q}{t}\Big\}\{q^r\}}{\{t\}\{q^{r-1}t^2\}}
\cdot \left(\prod_{j=1}^{r-1} \frac{\{q^j\}}{\{q^{j-1}t\}}\right)
\cdot
\frac{\Big\{\frac{A}{t^{s-r-1}}\Big\}\Big\{\frac{A}{t^{s+1}}\Big\}}
{\Big\{\frac{Aq^r}{t^{s}}\Big\}\Big\{\frac{Aq}{t^{s+1}}\Big\}}
& \approx \ \left(\frac{t}{q}\right)^{r+1} \cdot
\frac{\Big\{\frac{q}{t}\Big\}\{q^r\}}{\{t\}\{q^{r-1}t^2\}}
\cdot \left(\prod_{j=1}^{r-1} \frac{\{q^j\}}{\{q^{j-1}t\}}\right)
\label{alphar1.1p}
\ee
All other coefficients ${\cal B}_{([r,1],[1^p])}^{[r+1-m]}$
with $m=2,\ldots,r$ are not factorized corresponding to sums of skew Macdonald polynomials in the small/large $A$ limit.

Like it happened for $r=2$, the combination
\be
{\cal B}_{([r,1],[1^p])}^{[r]} M_{[r]} + {\cal B}_{([r,1],[1^p])}^{[r-1,1]} M_{[r-1,1]}
\ \approx \ \frac{t}{q}\cdot\frac{\{q\}}{\{t\}}\cdot M_{[r,1]/[1]} + O(A^{2})
\ee
reproduces the skew Macdonald
\be
M_{[r,1]/[1]} = M_{[r]} + \frac{\{t\}\{q^{r-1}\}\{q^rt\}\{q^{r-2}t^2\}}{\{q\}\{q^{r-2}t\}\{q^{r-1}t\}\{q^{r-1}t^2\}}
\cdot M_{[r-1,1]}
\label{Mr1/1}
\ee
in the limit of small/large $A$.

\bigskip

\framebox{\parbox{15cm}{The message that follows from these examples is very clear:
\begin{itemize}
\item[{\bf (a)}] what exists in general is a decomposition of the composite Macdonald polynomial
into the ordinary ones,
but, at finite $A$, it {\it can not} be reduced to a decomposition into
the skew Macdonald polynomials,
\item[{\bf (b)}] however, at $A^{\pm 1}\longrightarrow \infty$ such a skew Macdonald
decomposition exists at arbitrary $t$ and $q$,
\item[{\bf (c)}] but at $q\neq t$ this decomposition involves the terms with arbitrary sub-diagrams,
$R/\eta_1$ and $P/\eta_2$, not restricted by the constraint $\eta_2=\eta_1^\vee$, but only to $|\eta_2|=|\eta_1|$,
the restriction/correlation appears only at $t=q$.
\end{itemize}}}

\section{Conclusion}

In this paper, we discussed the definition of Kerov functions for
the composite representations $(R,S)$.
In the case of Schur functions, such a definition is provided by
the Koike formula (\ref{compSchur}) of \cite{Koike,Kanno,MMhopf}
and it is crucially important for the study of HOMFLY polynomials \cite{L8n8}.
However, its counterpart is not known even in the Macdonald case,
what is a serious obstacle for extending the results of \cite{L8n8} to superpolynomials.
The origin of difficulties is highlighted by the study in the general setting,
i.e. in the Kerov setting.
Our natural conjecture in this paper is that, in the Kerov case, the formula involves
a double sum over all diagrams, which precede $R$ and $P$ in the lexicographical ordering
(including smaller size ones),
and then some simplifications occur at the Macdonald and Schur levels:
\be
&\Ker_{(R,P)}^{(g)}[X] = {\cal X}^{l_{_{P^\vee}}} \cdot\sum_{\stackrel{R'\leq R}{S'\leq  S}}
 a_{R',S'}^{(N)}\cdot
\Ker^{(g)}_{[R']}[X]\cdot \Ker^{(g)}_{[S']}[\bar X]&
 \nn \\
  &\Biggr\downarrow \ \ \lefteqn{\begin{array}{c}{\rm Macdonald\ Locus,}\\{{\rm uniformization\ in}\ N}\end{array}}& \nn \\ \nn \\
 \!\!\!\!\!\!\!\!\!\!\!\!\!\!
&\Mac_{(R,P)} [X] = {\cal X}^{l_{_{P^\vee}}} \cdot\left(\Mac_{R}[X]\cdot \Mac_{P}[\bar X]
+\sum_{m=0}^{{\rm min}(|R|,|P|)-1}
\sum_{R',P'\vdash m}  {\cal B}_{R',P'}^{R,P}(A,q,t)\cdot
\Mac_{R'}[X]\cdot \Mac_{P'}[\bar X]
 \right) &
 \nn \\
& \Biggr\downarrow \ \ \lefteqn{\rm Schur\ Locus} &\nn \\ \nn \\
&{\rm Schur}_{(R,P)} [X] ={\cal X}^{l_{_{P^\vee}}} \cdot
\sum_\eta
(-1)^{|\eta|} {\rm Schur}_{R/\eta}[X]\cdot {\rm Schur}_{P/\eta^\vee}[\bar X]&
\ee
We illustrated this claim by a number of examples,
which are, in fact, rather tedious calculations.

\bigskip

The main subjects relevant to this story, which we only introduced,
and which should be developed much further seem to be:
\begin{itemize}

\item{Kerov functions for conjugate and composite representations
+ decomposition formulas for them }

\item{Universality of denominator functions + some formulas for them}

\item{The structure of Macdonald ideal}

\item{Uniformization of $N$-dependence at the Macdonald locus}

\end{itemize}

\noindent
Of special significance is search for other interesting loci where the Kerov functions acquire
special properties and thus provide yet unknown multi-parametric generalizations
of Macdonald polynomials, the obvious option are the
3-Macdonald polynomials
\cite{Zen3mac}, the hypothetical 3-Schur functions \cite{M3Schurs}
and/or the characters of the Pagoda \cite{Pag},
as well as generalized characters needed in tensor models \cite{IMMtenchar}.

\bigskip

We hope that this paper, together with \cite{MMker}, proves that
development of computer methods makes
the very difficult and long neglected topic of Kerov functions
available for efficient investigation,
and we can expect many new results in the near future.

\section*{Acknowledgements}

Our work is partly supported by the grants of the Foundation for the Advancement of Theoretical Physics
BASIS, by RFBR grants 19-01-00680 (A.Mir.), 19-02-00815 (A.Mor.), by the joint grants 19-51-50008-YaF (A.Mir.), 18-51-45010-Ind, 19-51-05015-Arm, 19-51-53014-GFEN.
We also acknowledge the hospitality of KITP and partial support by the National
Science Foundation under Grant No. NSF PHY-1748958 at a certain stage of this project.

\end{document}